\documentclass[]{aastex631}

\usepackage{rotating}

\shorttitle{ATLAS Phase Curves of KBOs, Centaurs, and JFCs}
\shortauthors{Dobson et al.}
\graphicspath{{./}{figuresA/}}

\begin{document}

\title{Phase Curves of Kuiper Belt Objects, Centaurs, and Jupiter Family Comets from the ATLAS Survey}

\author[0000-0002-1105-7980]{Matthew M. Dobson}
\affiliation{Astrophysics Research Centre, School of Mathematics and Physics, Queen's University Belfast, Belfast BT7 1NN, UK}

\author[0000-0003-4365-1455]{Megan E. Schwamb}
\affiliation{Astrophysics Research Centre, School of Mathematics and Physics, Queen's University Belfast, Belfast BT7 1NN, UK}

\author[0000-0001-8821-5927]{Susan D. Benecchi}
\affiliation{Planetary Science Institute, 1700 East Fort Lowell Rd., Suite 106, Tucson, AZ 85719, USA}

\author[0000-0002-3323-9304]{Anne J. Verbiscer}
\affiliation{Department of Astronomy, University of Virginia, P.O. Box 400325, Charlottesville, VA 22904-4325, USA}

\author[0000-0003-0250-9911]{Alan Fitzsimmons}
\affiliation{Astrophysics Research Centre, School of Mathematics and Physics, Queen's University Belfast, Belfast BT7 1NN, UK}

\author[0000-0002-5738-1612]{Luke J. Shingles}
\affiliation{GSI Helmholtzzentrum f\"{u}r Schwerionenforschung, Planckstraße 1, 64291 Darmstadt, Germany}
\affiliation{Astrophysics Research Centre, School of Mathematics and Physics, Queen's University Belfast, Belfast BT7 1NN, UK}

\author[0000-0002-7034-148X]{Larry Denneau}
\affiliation{University of Hawaii
Institute for Astronomy, 2680 Woodlawn Dr., Honolulu, HI 96822}

\author[0000-0003-3313-4921]{A. N. Heinze}
\affiliation{DiRAC Institute and the Department of Astronomy, University of Washington, 3910 15th Avenue NE, Seattle, WA 98195}

\author[0000-0001-9535-3199]{Ken W. Smith}
\affiliation{Astrophysics Research Centre, School of Mathematics and Physics, Queen's University Belfast, Belfast BT7 1NN, UK}

\author[0000-0003-2858-9657]{John L. Tonry}
\affiliation{University of Hawaii
Institute for Astronomy, 2680 Woodlawn Dr., Honolulu, HI 96822}

\author[0000-0003-1847-9008]{Henry Weiland}
\affiliation{University of Hawaii
Institute for Astronomy, 2680 Woodlawn Dr., Honolulu, HI 96822}

\author[0000-0002-1229-2499]{David. R. Young}
\affiliation{Astrophysics Research Centre, School of Mathematics and Physics, Queen's University Belfast, Belfast BT7 1NN, UK}

\begin{abstract}

The Kuiper belt objects, the Centaurs, and the Jupiter-family comets form an evolutionary continuum of small outer Solar System objects, and their study allows us to gain insight into the history and evolution of the Solar System. Broadband photometry can be used to measure their phase curves, allowing a first-order probe into the surface properties of these objects, though limited telescope time makes measuring accurate phase curves difficult. We make use of serendipitous broadband photometry from the long-baseline, high-cadence ATLAS survey to measure the phase curves for a sample of 18 Kuiper belt objects, Centaurs, and Jupiter-family comets with unprecedentedly large datasets. We find phase curves with previously reported negative slopes become positive with increased data and are thus due to insufficient sampling of the phase curve profile, and not a real physical effect. We search for correlations between phase curve parameters, finding no strong correlations between any parameter pair, consistent with the findings of previous studies. We search for instances of cometary activity in our sample, finding a previously reported outburst by Echeclus and a new epoch of increased activity by Chiron. Applying the main belt asteroid $HG_{1}G_{2}$ phase curve model to three Jupiter-family comets in our sample with large phase angle spans, we find their slope parameters imply surfaces more consistent with those of carbonaceous main belt asteroids than silicaceous ones. 

\end{abstract}


\section{Introduction}

Kuiper belt objects (KBOs) are small planetesimals residing beyond the orbit of Neptune and are the remnants of planet formation during the early history of the Solar System. 
Over 3700 KBOs\footnote{From Minor Planet Center: \url{https://www.minorplanetcenter.net}} have been discovered to date, and can be divided into sub-populations based on their present dynamics \citep[]{2006ssu..book..267D,2008ssbn.book...43G,2008ssbn.book...59K} and physical properties \citep[]{2007ApJ...659L..61S,2008ssbn.book..335B,2011ApJ...739L..60B}.
{Originating from the Kuiper belt population, the Centaurs are small icy objects that have diffused inwards onto giant planet-crossing orbits \citep[]{1997Icar..127...13L,1997Sci...276.1670D,2008ApJ...687..714V}}. Centaurs are likely precursors to certain populations of short-period comets, most notably the Jupiter-family comets (JFCs) \citep[]{2003AJ....126.3122T,FERNANDEZ20131138,2013DPS....4550806B,2019ApJ...883L..25S}, whose orbital dynamics are governed predominantly by Jupiter's gravitational influence \citep[]{1973A&A....24..107V,1987PAICz..67...21C,1994Icar..108...18L,1996ASPC..107..173L,2008ssbn.book...43G,2019ApJ...883L..25S} and whose perihelion distances bring them sufficiently close to the Sun for water ice to sublimate {(\citealp{2017PASP..129c1001W} and references therein)}. Studying these populations of primordial objects can reveal information about the conditions, dynamics and evolutionary processes of the Solar System, both in its early history and its present state.

Broadband photometry remains the fastest method of characterizing large numbers of these objects, 
as they are generally too faint for spectroscopy. 
Such photometry can be used to analyse their phase curves: the change in an object's distance-corrected brightness when observed at different solar phase angles $\alpha$, the angle between the Sun and the observer as viewed from the object. 
The phase curve of an atmosphereless object is determined by its surface reflectivity, itself a complex function of its surface properties. Thus, analysis of an object’s phase curve can allow an explorative probe into its overall surface properties \citep[]{1963AJ.....68..279H,1966AJ.....71S.386H,1968Sci...159...76H,1981JGR....86.3039H,1984Icar...59...41H,1986Icar...67..264H,2002Icar..157..523H,2008Icar..195..918H,2012Icar..221.1079H,2021Icar..35414105H,1989aste.conf..524B,2000Icar..147..545N,2008ssbn.book..115B,1998ASSL..227..157V,2013ASSL..356...47V}.

The large heliocentric distances of KBOs and many Centaurs limit Earth-based observations of these objects to a narrow range of small phase angles ($\alpha \lesssim 2$ deg), requiring in-situ measurements from spacecraft for high phase angles. \citet{2019AJ....158..123V} and \citet{2022PSJ.....3...95V} obtained phase curves of KBOs spanning several dynamical sub-populations by combining data from both Earth-based observations and \emph{NASA's} \emph{New Horizons} spacecraft, extending the phase angle coverage beyond that achievable from Earth. The phase curves of the KBOs analysed were found to exhibit similarities to those of other populations of small Solar System bodies, including JFCs and giant planet satellites theorized to be captured KBOs \citep{2019AJ....158..123V}. {Additionally, KBOs with high-albedo surfaces dominated by volatile ices were found to exhibit shallow phase curve slopes, with steeper slopes being a feature of objects with lower-albedo surfaces \citep{2015aste.book..129L,2018MNRAS.481.1848A,2022PSJ.....3...95V}}. Without such measurements, the small phase angle range observable from Earth impairs the ability of 
surface reflectance models to accurately fit the data. However, {beyond the very small range of phase angles where the opposition surge - the sudden increase in an object's brightness as $\alpha \rightarrow 0$ deg - is significant,} Centaur and KBO phase curves typically exhibit approximately linear profiles across phase angles observable from Earth \citep{1984AJ.....89.1759B,1994Icar..108..200T,1997Icar..125..233B,2002AJ....124.1757S,2002Icar..160...52S,2006ApJ...639.1238R,2007AJ....133...26R,2007AJ....134..787S,2008ssbn.book..115B,2009AJ....137..129S,2013ASSL..356...47V}, with a slope quantified by the linear phase coefficient, $\beta$. Fitting a linear function to the phase curves generated using Earth-based observations has thus allowed comparison of phase curves between individual objects and populations.

\citet{2007AJ....133...26R} and \citet{2009AJ....137..129S} found phase curves of KBOs and Centaurs to exhibit approximately linear trends when observed from Earth. These phase curves ranged from flat to steep in profile, with varying phase coefficient values between different photometric filters. The largest KBOs exhibited small phase coefficients ($\beta < 0.1$ mag deg$^{-1}$), suggesting such a property may be indicative of objects with high-albedo volatile-rich surfaces, in accordance with both the findings of \citet{2022PSJ.....3...95V} and theoretical models of surface reflectivity. Several Centaurs also exhibited similarly flat phase curves, implying their surfaces are dominated by organic compounds. This accords with the findings of \citet{2002AJ....124.1757S}, laboratory measurements of both high- and low-albedo materials \citep{2002Icar..159..396S} and the improbability of volatile retention by Centaurs due to their smaller heliocentric distances.
\citet{2009AJ....137..129S} reported correlations between an object's opposition surge with its color index and geometric albedo. Both \citet{2009AJ....137..129S} and \citet{2019AJ....158..123V} find a distinction between the scattering properties of outer Solar System satellites and those of the Centaur and KBO populations. 
The studies of \citet[]{2016A&A...586A.155A,2018MNRAS.481.1848A,2019MNRAS.488.3035A} report several phase coefficients for Centaurs and KBOs whose values were significantly negative, implying unphysical dimming at small phase angles which cannot be explained by surface reflectivity models. These studies also 
reported a strong correlation between the relative phase coefficient between filters (defined by \citet{2019MNRAS.488.3035A} as $\Delta\beta \equiv \beta_V - \beta_R$) and an object's color index ($H_{V}- H_{R}$){, in addition to a correlation between relative phase coefficient and albedo for objects with absolute magnitudes $H_{V}\geq4.5$.} 
\citet{2017MNRAS.471.2974K} and \citet{2018MNRAS.479.4665K} measured shallow phase coefficients of $\beta < 0.1$ mag deg$^{-1}$ for 14 JFCs, and reported a possible correlation of increasing geometric albedo with steeper phase coefficient.

Phase curve measurements require high-cadence observations over a long baseline to sample the full range of phase angles visible from Earth. Such ideal measurements are difficult to achieve in practice, due to the limited availability of telescope time, in addition to the challenges of ground-based observations. Combining data of an object from multiple telescopes at different sites, as well as from literature studies, can help augment otherwise sparse datasets \citep[]{2016A&A...586A.155A,2018MNRAS.481.1848A,2019MNRAS.488.3035A}. However, this could introduce systematic errors into the dataset due to heterogeneous data reduction pipelines and differing broadband filter wavelength ranges, as well as potential long timespans between observations. To sample the full range while ensuring all data come from a homogeneous photometric system and data reduction pipeline would require observations with a telescope that performs regular-cadence observations of that object. Such an opportunity is afforded by the Asteroid Terrestrial-impact Last Alert System (ATLAS) survey \citep[]{2018PASP..130f4505T,2018ApJ...867..105T}, which has accumulated large datasets of serendipitous observations of many KBOs, Centaurs, and JFCs.

In this paper, we make use of the high-cadence, long baseline photometry obtained by the ATLAS survey to generate phase curves for a sample of KBOs, Centaurs, and JFCs visible to the ATLAS telescopes in two broadband filters. With the advantage of a larger dataset than previous studies across a single baseline, we measure the phase coefficients of these objects, search for correlations between these and other object parameters, and look for outbursts of cometary activity across the survey's near-continuous, 6-year observation baseline. This paper is structured as follows. 
In section 2, we describe the ATLAS survey. Section 3 details the sample of objects selected for this study. In section 4, we present details about the photometry obtained by and used in this study. Section 5 details the data analysis methods, while the results from this analysis and our discussion of these are presented in Sections 6. Finally, in Section 7, we draw our conclusions from our analysis.

\section{ATLAS}

The data used in this study originate from observations performed by the ATLAS wide-field survey \citep{2018PASP..130f4505T}, whose primary mission is to detect near-Earth asteroids that may pose an impact hazard to Earth.
Since its first observations in 2016, ATLAS has serendipitously observed a sample of bright KBOs, Centaurs, and JFCs, with the long baseline and high cadence of observations providing ample sampling of the phase curves of these objects. This eliminates the need to include additional data from external sources, thereby reducing systematic errors that could arise from combining data obtained via heterogeneous methods of image processing and data analysis. ATLAS's long baseline has been previously utilized by studies analysing other populations of small Solar System objects, such as asteroids \citep[]{2020ApJS..247...13E,2021Icar..35414094M} and Jupiter Trojans \citep{2021PSJ.....2....6M}.

ATLAS presently consists of four 0.5-m Schmidt telescopes, two of which are located at Hawai'i (sites at Haleakalā and Maunaloa), with another two installed in 2022 at sites in Chile and South Africa. The ATLAS telescopes regularly observe the sky outside $60$ deg of the Sun to a limiting magnitude of approximately $19.5$ mag in two non-standard wide-band filters - \textit{cyan} (\textit{c}, spanning 420-650 nm) and \textit{orange} (\textit{o}, spanning 560-820 nm). When weather conditions permit, four 30s exposures are taken per pointing over a 1 hour interval. Each exposure covers 28.9 deg$^{2}$ field of view \citep{2018PASP..130f4505T,2018ApJ...867..105T}, with the two telescopes of ATLAS installed at the time of this study covering the entire accessible sky over approximately 2 days. Further details of the ATLAS system and data reduction pipeline are described in \citet{2018PASP..130f4505T,2018ApJ...867..105T} and \citet{2020PASP..132h5002S}. For this study, we make use of data from the two ATLAS telescopes at the Hawai'ian sites of Haleakalā and Maunaloa, as the southern hemisphere telescopes had only recently been installed at the time of this study.

\section{Sample of Objects}

We select objects from the KBO, Centaur and JFC populations that may have been observable by ATLAS, downloading the 2021 catalog of objects from JPL Horizons\footnote{https://ssd.jpl.nasa.gov/ - retrieved on 2021 March 25}.
These objects are selected according to the following criteria:

\begin{enumerate}
    \item The semimajor axis of each object must lie beyond the aphelion of Jupiter ($a > 5.5$ au).\item For ${\geq}1$ month during the ATLAS baseline, the object must be brighter than $19.5$ mag (ATLAS limiting magnitude) to ensure detection.\item The objects' predicted positional uncertainties had to be ${<}1^{\prime\prime}$ (equating to ${\sim}$0.5 pixels on ATLAS images), to ensure proper identification on the ATLAS images;
    \item The ATLAS observations of each object had to span ${>}$50\% of the total observable phase angle range during the time that ATLAS has been observing, to ensure adequate sampling of at least half the observable span of the phase curve;
    \item Each object had to have at least 50 observations recorded in both \textit{c} and \textit{o} filters, to ensure adequate sampling.\end{enumerate}

Upon selecting our sample of objects, we subsequently divide them into their dynamical classes. We define KBOs as small outer Solar System objects whose semimajor axes lie beyond that of Neptune ($a > a_N$). The KBOs in our sample come from several dynamically hot sub-populations of the Kuiper belt using the definitions of \citet{2008ssbn.book...43G}: 3 objects from the resonant population (Pluto, Orcus, Huya), 1 from the detached TNO population (Eris), and 3 from the hot classical Kuiper belt (Haumea, Quaoar, Makemake). 
Furthermore, 6 of our 7 KBOs, with absolute magnitudes $H_{V}<3$, differ significantly from the remainder of the Kuiper belt. These Pluto-sized objects have greater masses, allowing retention of high albedo volatiles on their surfaces, augmenting their absolute magnitude such that they deviate from the brightness distribution of the Kuiper belt \citep[]{2004AJ....128.1364B,2007DPS....39.4910S,2008ssbn.book..335B,2012AREPS..40..467B}.

At present, there is no agreed upon definition for the Centaur population, nor is there a consensus distinction between the Centaurs and the related JFC population \citep[]{1997Icar..127...13L,2008ssbn.book...43G,2009AJ....137.4296J,2017MNRAS.471.2974K,2020ApJ...892L..38C,2021PSJ.....2..155L,2021Icar..35814201R}. For our study, we use a dynamical definition to separate these populations not based on past histories of cometary activity. We define and distinguish between the Centaur and JFC populations by their semimajor axes, perihelia, and the effect of Jupiter on their dynamics, the latter of which can be quantified by the Tisserand parameter with respect to Jupiter, $T_J$, defined as:

\begin{equation}
    T_J \equiv \frac{a_J}{a} + 2\cos i \sqrt{\frac{a}{a_J}(1-e^2)}
\end{equation}

\hfill
\break
where $a$ and $a_J$ are the semimajor axes of the object and Jupiter respectively, $e$ the object's eccentricity and $i$ the object's inclination. We construct our Centaur definition based on those of \citet{2008ssbn.book...43G} and \citet{2009AJ....137.4296J}, classifying Centaurs as small Solar System objects not in 1:1 mean motion resonance with any planet, on orbits with semimajor axes $a$ and perihelia $q$ between the semimajor axes of Jupiter and Neptune ($a_J < a,q < a_N$), and Tisserand parameters with respect to Jupiter $T_J > 3.05$, this  threshold value accounting for Jupiter's non-zero eccentricity and the resulting effect on small object dynamics \citep{2008ssbn.book...43G}. We accordingly define Jupiter-family comets (JFCs) as small Solar System objects not in 1:1 mean motion resonance with a planet, whose perihelia lie interior to the orbit of Jupiter ($q < a_J$) and with Tisserand parameters with respect to Jupiter of $2 < T_J \leq 3.05$. 
We classify any objects in our sample with $a < a_N$ that do not satisfy the definitions for Centaurs or JFCs as \emph{Transition Objects}. Table \ref{table_Definitions} lists the objects in our sample with semimajor axes $a < a_N$, and their dynamical classification according to the definitions of Centaurs and JFCs from \citet{2008ssbn.book...43G}, \citet{2009AJ....137.4296J} and \citet{2019ApJ...883L..25S}, in addition to their classification in this study.

\begin{deluxetable*}{lcccr}
\tablecaption{Classification of objects used in this study with semimajor axes $5.5$ au $< a < 30.1$ au, according to the definitions of \citet{2008ssbn.book...43G}, \citet{2009AJ....137.4296J} and \citet{2019ApJ...883L..25S}.  \label{table_Definitions}}
\tablewidth{0pt}
\tablehead{
\colhead{Name} & \colhead{\citet{2008ssbn.book...43G}} & \colhead{\citet{2009AJ....137.4296J}}  & \colhead{\citet{2019ApJ...883L..25S}} & \colhead{This study}
}
\startdata
   (944) Hidalgo&JFC&JFC&Neither&JFC\\
  (2060) 95P/Chiron&Centaur&Centaur&Centaur&Centaur\\
 (10199) Chariklo&Centaur&Centaur&Centaur&Centaur\\
 (37117) Narcissus&JFC&JFC&Neither&JFC\\
 (54598) Bienor&Centaur&Centaur&Centaur&Centaur\\
 (60558) 174P/Echeclus&JFC&Centaur&Centaur&Transition Object\\
(347449) 2012 TW236&JFC&JFC&Neither&JFC\\
(349933) 2009 YF7&JFC&Centaur&Centaur&Transition Object\\
(459865) 2013 XZ8&Centaur&Centaur&Centaur&Centaur\\
(501585) 2014 QA43&JFC&JFC&Neither&JFC\\
       2016 ND21&JFC&JFC&Neither&JFC\\
\enddata
\end{deluxetable*}

\begin{figure}
\centering
\includegraphics[width=\textwidth]{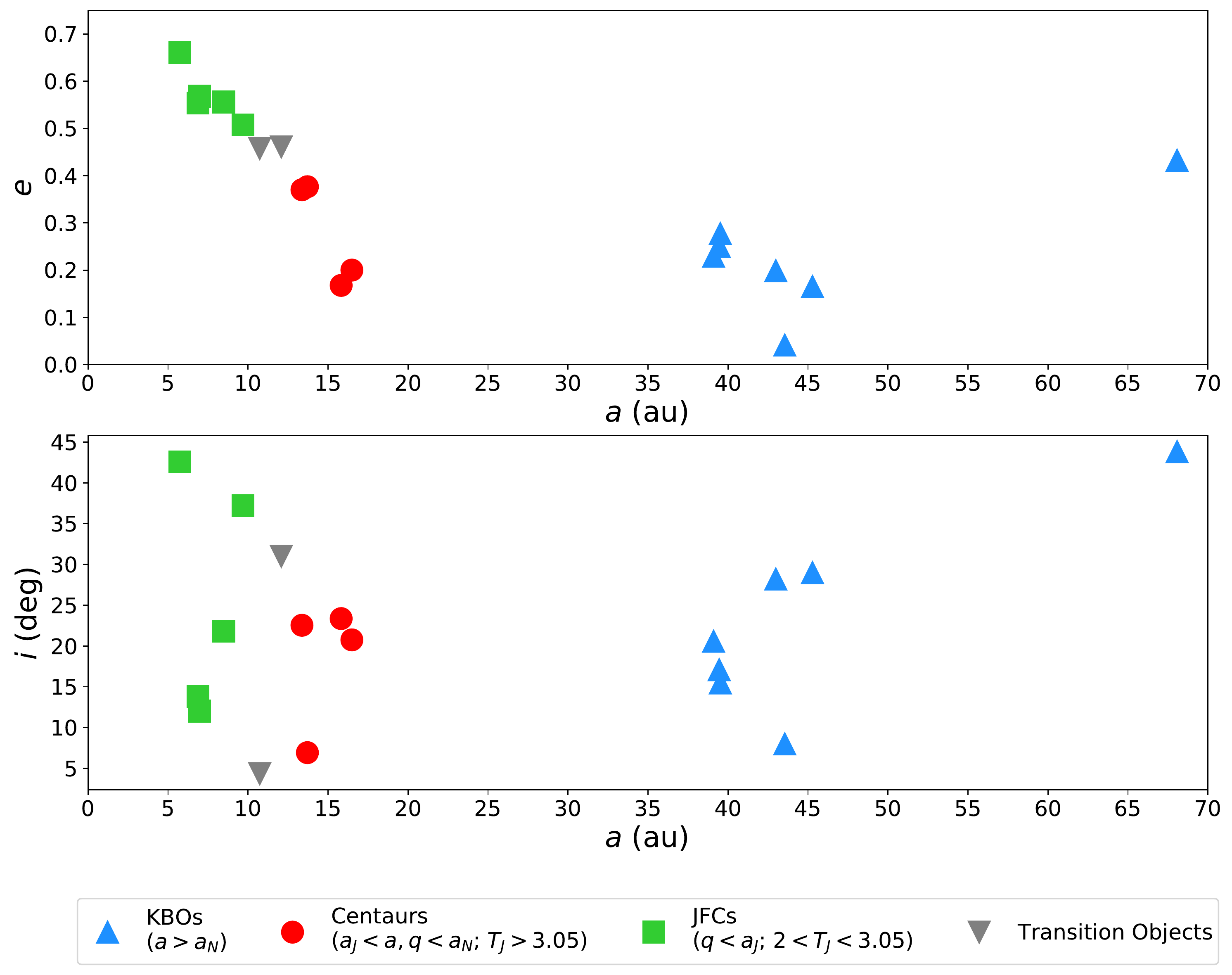}
\caption{Orbital distribution diagram showing semimajor axes ($a$), orbital eccentricities ($e$) and orbital inclinations ($i$) of the objects used in this study. Data taken from JPL Horizons.}
\label{fig1}
\end{figure}

\begin{figure}
\centering
\includegraphics[width=0.7\columnwidth]{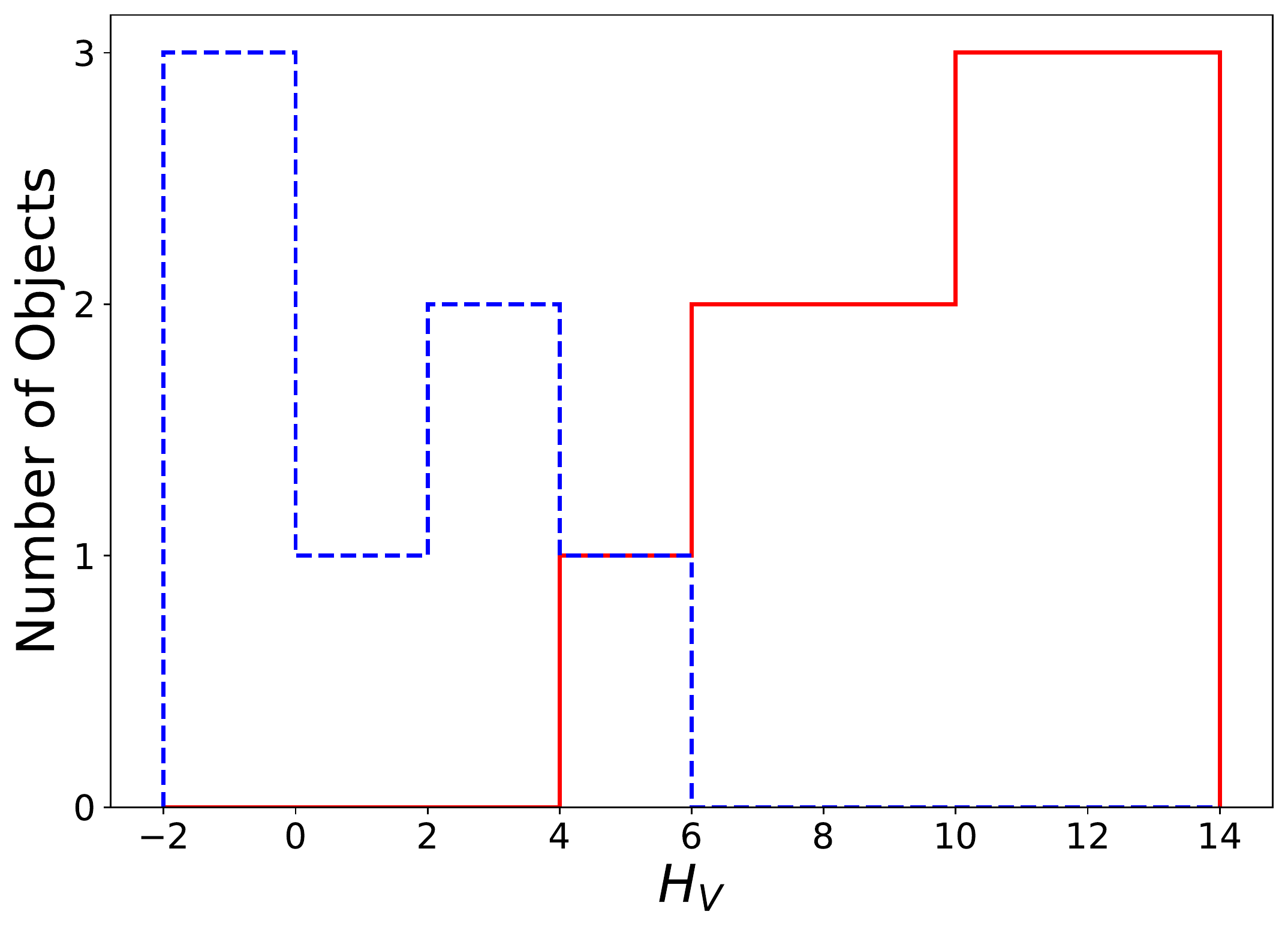}
\caption{Histogram of \emph{V}-band absolute magnitudes from JPL Horizons of the sample of objects with semimajor axis less than Neptune's $a < a_N$ (solid red) and greater than Neptune's $a > a_N$ (dashed blue) used in this study. Bin-size = 2 mag.}
\label{fig2}
\end{figure}

\begin{figure}
\centering
\includegraphics[width=\textwidth]{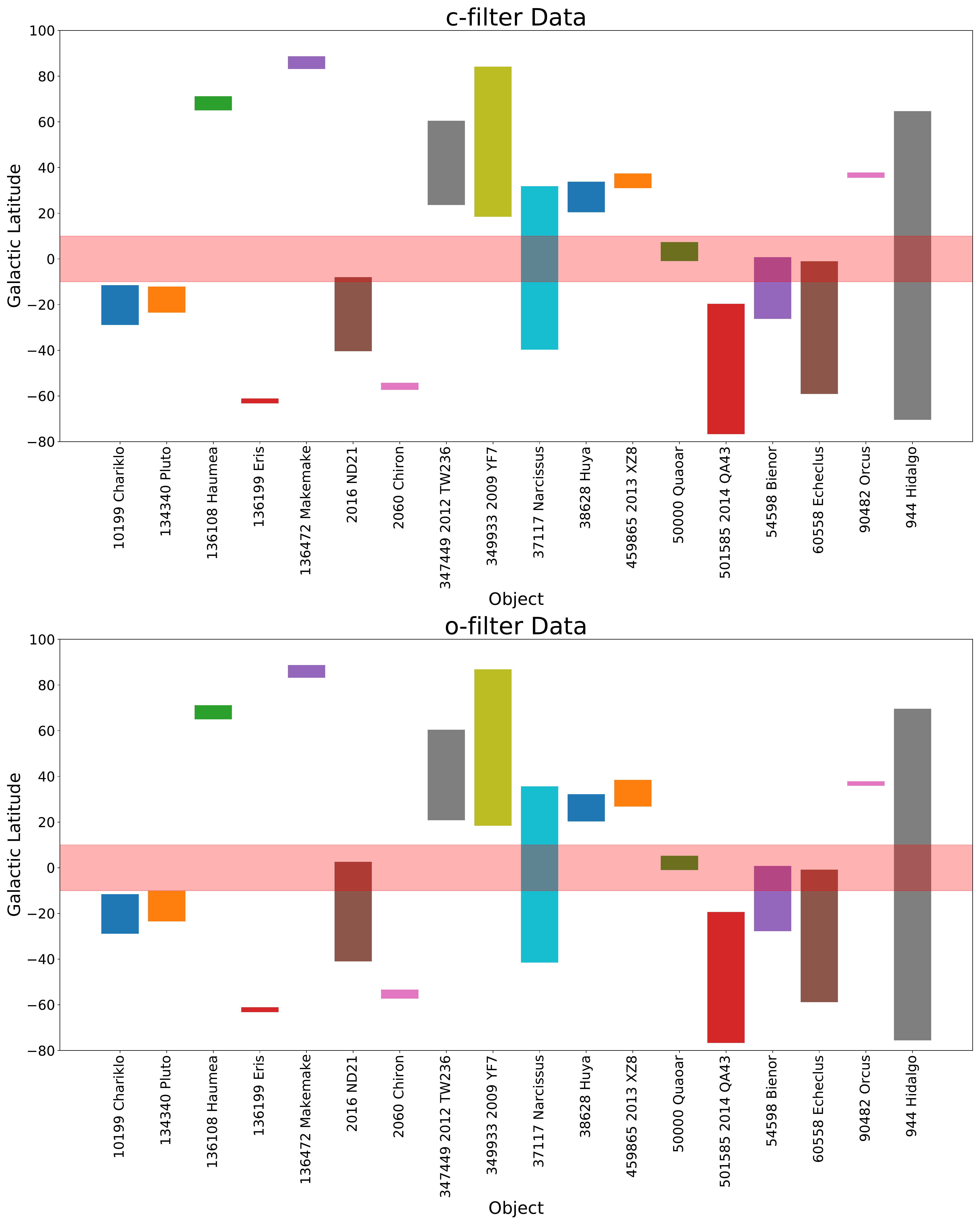}
\caption{Bar plot of galactic latitude of observations for each object in ATLAS $c$ (upper plot) and $o$ (lower plot). The red bar denotes the sky region spanning $\pm10$ deg from the galactic equator.}
\label{SampleGalacticLatitude}
\end{figure}

Our sample consists of 18 objects (7 KBOs, 4 Centaurs, 5 JFCs, and 2 Transition Objects), whose orbital parameters are listed in Table \ref{TableSampleInvariantData}, along with their estimated V-band absolute magnitudes, and, if known, their published rotational {lightcurve} periods and amplitudes. The orbital distribution of our sample is shown in Figure \ref{fig1} and its $H$ distribution is plotted in Figure \ref{fig2}. 
There are a median of 181 and 463 ATLAS observations per object in the \emph{c} and \emph{o} filters, respectively. 
Across all objects in this study, the ATLAS data cover a median of  93\% (\emph{c}) and 95\% (\emph{o}) of the total observable phase angle range from Earth across the ATLAS baseline.
These objects span apparent magnitudes $14.2 < c < 19.8$ and $13.8 < o < 19.7$ in each ATLAS photometric filter. 
The galactic latitudes spanned by our sample of objects during the ATLAS baseline are shown in Figure \ref{SampleGalacticLatitude}, showing many objects in our sample cross into the galactic plane where stellar crowding and the resulting background flux contamination must be dealt with.

\begin{deluxetable*}{lccccccccccr}
\tablecaption{Heliocentric orbital elements, absolute magnitudes and rotation periods of KBOs, Centaurs, JFCs and Transition Objects visible to ATLAS survey used in this study. $a$ = semimajor axis; $e$ = eccentricity; $i$ = inclination; $q$ = perihelion; $Q$ = aphelion; $T_J$ = Tisserand parameter (with respect to Jupiter); $H$ = (\emph{V}-band) absolute magnitude; $P$ = rotation period; $\Delta M$ = peak-to-peak rotational {lightcurve} amplitude. 
\label{TableSampleInvariantData}
}
\tabletypesize{\scriptsize}
\tablehead{
\colhead{Name} & \colhead{$a$} & 
\colhead{$e$} & \colhead{$i$} & 
\colhead{$q$} & \colhead{$Q$} & 
\colhead{$T_J$} & \colhead{$H$} & 
\colhead{$P$} & \colhead{$\Delta M$} & \colhead{Source for $P$} & \colhead{Dynamical Class}\\ 
\colhead{} & \colhead{(au)} & \colhead{} & \colhead{(deg)} & 
\colhead{(au)} & \colhead{(au)} & \colhead{} &
\colhead{(mag)} & \colhead{(hrs)} & \colhead{(mag)} & \colhead{and $\Delta M$} & \colhead{}
} 
\startdata
(136199) Eris&67.94&0.43&43.97&38.41&97.51&4.780&-1.11&...&...&...&KBO (detached) \\
(136108) Haumea&43.12&0.2&28.21&34.65&51.54&5.086&0.25&3.915341&0.28&SS17&KBO (hot classical)\\
 (38628) Huya&39.69&0.28&15.47&28.55&50.49&5.235&4.86&5.28&0.02 $\pm$ 0.01&T14&KBO (3:2 resonant)\\
& & & & & & & & 4.45 & 0.1 & G16 \\
& & & & & & & & 6.75 & $<0.1$ & O03 \\
(136472) Makemake&45.37&0.16&28.99&37.98&52.77&5.204&-0.13&22.83&0.032 $\pm$ 0.005&H19&KBO (hot classical)\\
&&&&&&&&&0.0286 $\pm$ 0.0016&HL09 \\
 (90482) Orcus&39.14&0.23&20.59&30.21&48.06&5.129&2.3&10.5&${\sim}0.05$&G16&KBO (3:2 resonant)\\
&&&&&&&&&0.18 $\pm$ 0.08&R06 \\
&&&&&&&&&0.03-0.18&D10 \\
&&&&&&&&&0.06 $\pm$ 0.04&O11 \\
(134340) Pluto&39.45&0.25&17.09&29.57&49.32&5.228&-0.4&153.29352&0.19&B18&KBO (3:2 resonant)\\ 
&&&&&&&&&(Pluto+Charon)&&\\
 (50000) Quaoar&43.66&0.04&7.99&41.85&45.35&5.844&2.51&8.8394&0.17 $\pm$ 0.04&O03 &KBO (hot classical)\\
&&&&&&&&&0.17&FB10 \\
\hline
(459865) 2013 XZ8&13.42&0.37&22.53&8.42&18.41&3.140&9.69&...&...&...&Centaur \\
 (54598) Bienor&16.46&0.2&20.75&13.18&19.71&3.578&7.55&9.14&0.7&O02&Centaur \\
&&&&&&&&&0.75 $\pm$ 0.09&O03 \\
&&&&&&&&&0.34 $\pm$ 0.08&R06 \\
&&&&&&&&&0.088 $\pm$ 0.008 (2014)&FV17 \\
&&&&&&&&&0.082 $\pm$ 0.007 (2015)&FV17 \\
&&&&&&&&&0.10 $\pm$ 0.02 (2016)&FV17 \\
 (10199) Chariklo&15.84&0.17&23.35&13.17&18.52&3.485&6.54&7.004&0.11-0.13&F14&Centaur \\
&&&&&&&&&0.06 ± 0.02&L17 \\
  95P/(2060) Chiron&13.69&0.38&6.93&8.52&18.87&3.365&5.84&5.918&0.088 ± 0.003&LJ90&Centaur \\
&&&&&&&&&0.07 ± 0.02&D91 \\
&&&&&&&&&0.04 ± 0.005&MB93 \\
&&&&&&&&&0.06 ± 0.01&L97 \\
\hline
(347449) 2012 TW236&6.97&0.57&11.96&3.0&10.94&2.610&12.3&...&...&...&JFC \\
(501585) 2014 QA43&9.66&0.51&37.21&4.76&14.54&2.409&11.4&11.558&0.236&S18&JFC \\ 
      2016 ND21&8.47&0.56&21.83&3.76&13.16&2.584&12.3&17.53&0.31&H19&JFC \\
  (944) Hidalgo&5.73&0.66&42.54&1.95&9.52&2.068&10.69&10.0630&0.48&W97&JFC \\
 (37117) Narcissus&6.88&0.55&13.8&3.07&10.68&2.616&13.29&...&...&...&JFC \\
\hline
(349933) 2009 YF7&12.11&0.46&30.99&6.51&17.72&2.749&10.91&...&...&...&Transition Object \\
 174P/(60558) Echeclus&10.71&0.46&4.34&5.81&15.60&3.031&9.17&26.802&0.24 ± 0.06 (R-filter)&R05&Transition Object \\
 &&&&&&&&&0.36 ± 0.09 (V-filter)&R05 \\
\hline
\enddata
\tablerefs{Heliocentric orbital elements and absolute magnitude values obtained from JPL Horizons. Rotational period/amplitude sources: B18: \citet{2018Icar..314..265B}, SS17: \citet{2017AA...604A..95S}, O03: \citet{2003AA...409L..13O}, FB10: \citet{2010ApJ...714.1547F}, LJ90: \citet{1990AJ....100..913L}, D91: \citet{1991MNRAS.250..115D}, MB93: \citet{1993Icar..104..234M}, L97: \citet{1997PSS...45.1607L}, FV17: \citet{2017MNRAS.466.4147F}, O02: \citet{2002AA...388..661O}, D10: \citet{2010AA...520A..40D}, O11: \citet{2011AA...525A..31O}, G16: \citet{2016ApSS.361..212G}, F14: \citet{2014AA...568L..11F}, L17: \citet{2017AJ....154..159L}, R05: \citet{2015DPS....4721102R}, HL09: \citet{2009AJ....138..428H}, H19: \citet{2019AA...625A..46H}, T14: \citet{2014AA...569A...3T}, S18: Safrit (2018), 
W97: \citet{1997Icar..126..395W}.}
\end{deluxetable*}

\section{Data and Photometry}

We utilize observations made by the two ATLAS telescopes at Haleakalā and Maunaloa.
Data for our sample was obtained using the ATLAS Forced Photometry Server\footnote{https://fallingstar-data.com/forcedphot/} {\citep{2021TNSAN...7....1S}}, which fits a point-spread function (PSF) at an object's position on the image of an observation as predicted from its orbital ephemeris cataloged by the Minor Planet Center{, calculating the AB magnitude of the flux at that point}. The ATLAS data reduction includes an image subtraction pipeline, which matches a high signal-to-noise reference image of the static sky (``wallpaper'') to the location of a given observation's image, and subtracts the wallpaper flux from that image, creating a ``difference image'' containing only the flux not previously present
\citep[]{2018PASP..130f4505T,2018ApJ...867..105T}. We chose to use difference images to reduce the effect of contaminating background stars and galaxies on the brightness measurements of each object, which can become significant if an object passes in front of or near a background star on the sky. We select all ATLAS data up to and including the most recent observations at the time of selection (2021 June 30), allowing us to obtain sufficient photometry to generate phase curves for all 18 objects in our sample in both \emph{c} and \emph{o} filters. We deem any observation a detection if its apparent magnitude is brighter than 1) the $5\sigma$ limiting magnitude of the image (to ensure the observation was of good quality); and 2) the $3\sigma$ upper magnitude limit derived from the flux uncertainty (ensuring an observed object could be detected on the image). For every detection that satisfies these criteria, we correct their {Modified} Julian Dates (MJDs) for light travel time. {The ATLAS photometry used in this analysis for all objects in our sample are listed in Table \ref{ATLASTableAllData}.}

\begin{deluxetable*}{lccccccr}
\tablecaption{ATLAS photometry for each object.\label{ATLASTableAllData}}
\tablewidth{0pt}
\tablehead{
\colhead{Name} & \colhead{MJD} & \colhead{Phase}  & \colhead{Apparent Magnitude} & \colhead{Apparent Magnitude Uncertainty} & \colhead{Filter} & \colhead{Reduced Magnitude} & \colhead{Flags}\\
\colhead{} & \colhead{(light travel time-corrected)} & \colhead{(deg)}  & \colhead{} & \colhead{} & \colhead{} & \colhead{} & \colhead{}
}
{
\startdata
10199 Chariklo&57511.477108&3.1056&18.516&0.17&c&6.745&G\\
10199 Chariklo&57891.519318&2.689&18.771&0.151&o&6.935&G\\
10199 Chariklo&57895.513409&2.5112&18.579&0.115&c&6.749&G\\
10199 Chariklo&57903.490523&2.1231&19.009&0.124&c&7.189&G\\
10199 Chariklo&57903.507429&2.1222&19.243&0.129&c&7.423&G\\
\enddata
}
\tablecomments{This table is published in its entirety in the machine-readable format. A portion is shown here for guidance regarding its form and content.}
\tablecomments{Key for flags: G - data used to measure phase curve parameters; C - cometary activity flagged by detection algorithm; D - data not flagged by algorithm yet obtained during extended cometary activity; O - opposition effect.}
\end{deluxetable*}

\section{Data Analysis}\label{DataAnalysis}

We detail the methods used to fit the phase curves for our sample in both ATLAS $c$ and $o$ filters, search for rotational {lightcurves} in the ATLAS data, detect for epochs of cometary activity, and re-fit the data corrected for these effects. 

\subsection{Generating Phase Curves}

All ATLAS apparent magnitude measurements were transformed to reduced magnitudes to remove distance effects. The reduced magnitude $M$ is defined as:
\begin{equation}
    M = m - 5\log_{10}{(r\Delta)}
\end{equation}
where $m$ is the apparent magnitude, and $r$ and $\Delta$ are the heliocentric and geocentric distances of the object at the time of each observation, respectively. 
Distances and solar phase angles for each observation of a given object were obtained from JPL Horizons. The uncertainty in reduced magnitude was taken to be that of the measured apparent magnitude, due to the comparatively negligible uncertainties of the geocentric and heliocentric distances. 
{Datapoints with uncertainties larger than 85\% of all datapoints were removed. These observations were likely taken during poor observing conditions, causing their large uncertainties.}
Each object's dataset was {then} clipped to remove outlying points beyond 3-sigma from the median reduced magnitude to exclude measurements from contaminating flux, poor image subtraction, or poor calibration.
All data in a given filter were treated as a single bin of phase angle unless observations extended to phase angles ${>}20$ deg or ${<}0.1$ deg, where phase curve profiles become non-linear, in which cases data were split into two or four bins of phase angle, respectively. All data in each bin were sigma-clipped separately until convergence.

\subsection{Phase Curve Fitting}

\begin{figure}
\centering
\includegraphics[width=\columnwidth]{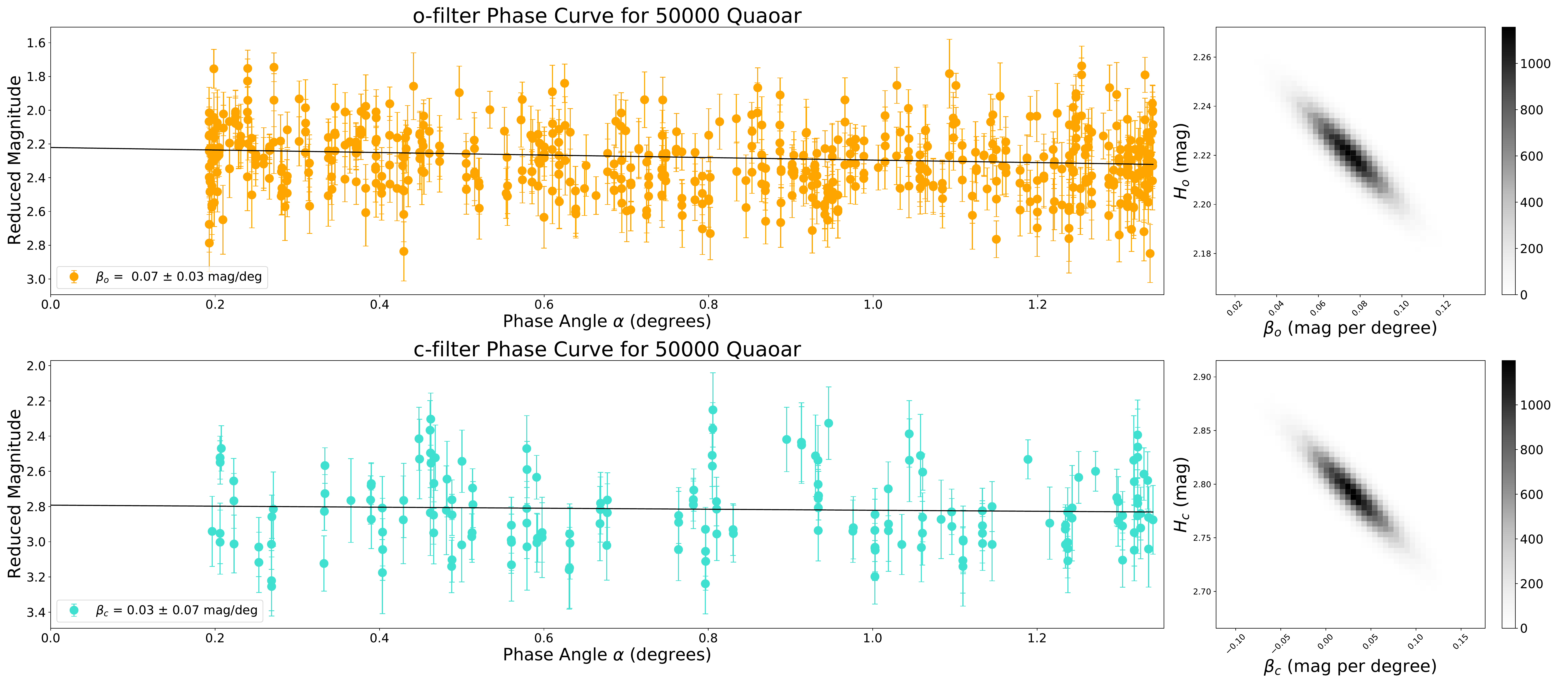}
\caption{Linear fit to phase curve of (50000) Quaoar in $c$ and $o$
    filters, with 2D histograms showing density plot of error
    distributions of linear phase coefficient and absolute magnitude.
    {Best fit function extended to $\alpha = 0$ deg for ease of
    viewing the extrapolated absolute magnitude value and any
    potential opposition surge.}
    The complete figure set (18 images) is available in the online published version of the paper.}
\label{PhCExampleQuaoar}
\end{figure}

{Beyond very small phase angles ($\alpha \sim 0.1$ deg, \citealp{2008ssbn.book..115B}) where the opposition effect is significant,} KBO, Centaur, and JFC phase curves exhibit approximately linear profiles when observed from Earth.  We apply a weighted linear fit to the phase curve of each object in each ATLAS filter, using the \emph{polyfit} function from the \emph{NumPy} {Python} package \citep[]{2011CSE....13b..22V,harris2020array} to fit a 1st-degree polynomial to the reduced magnitudes $M(\alpha)$ according to 
\begin{equation}
    M(\alpha) = H + \alpha\beta
\end{equation}
where $\alpha$ is the solar phase angle of the object at the time of measurement, $\beta$ is the linear phase coefficient, and $H$ is the absolute magnitude of the object. We consider only data at phase angles $\alpha \geq 0.1$ deg to remove any influence of an opposition surge on the resulting $\beta$ value. Each data point was weighted by the inverse of its magnitude uncertainty. The best fit values of both $\beta$ and $H$ and their associated uncertainties were computed via Monte Carlo simulation. We generate $10^5$ synthetic phase curves with synthetic reduced magnitude measurements $M_{i,synthetic}$ calculated according to:
\begin{equation}
    M_{i,synthetic} = M_i + {rand[0,\delta M_{i}]}
\end{equation}
where $rand[0,\delta M_{i}]$ is {a random number generated from a normal distribution centred at zero of standard deviation equal to $\delta M_{i}$, the uncertainty of each reduced magnitude measurement.}
The best fit $\beta$ and $H$ values were taken to be the medians of the distributions of all $10^{5}$ results for each parameter. These distributions tended to be approximately Gaussian in profile, thus we took the associated uncertainties of these parameters to be half of the range that includes the central 95\% of their corresponding distributions, yielding $2\sigma$ uncertainty values. 
Figure \ref{PhCExampleQuaoar} shows an example for the KBO Quaoar of an ATLAS phase curve of our sample and the associated uncertainty distributions.
For objects for which changes in brightness due to their periodic {lightcurve} or outbursts of cometary activity can be identified (applicable to 5 objects in total), we correct their datasets for these effects (see Sections \ref{LightCurveCorr} and \ref{CometActivityCorr} for {lightcurve} and cometary activity correction, respectively) and reapply the phase curve fitting algorithm to the corrected data.

\subsection{{Lightcurve} Corrections} \label{LightCurveCorr}

We aim to correct for the effect of rotational modulation on the phase curves wherever possible by searching for periodicities corresponding to the period of the rotational {lightcurve} of an object for all ATLAS data in our sample. For most objects in our sample, the uncertainty in the reduced magnitude measurements exceeds the literature peak-to-peak amplitude of the rotational {lightcurve}. This obscures any rotational modulation, thus precluding any correction of this effect on the phase curve of the object. However, for some objects in our sample we observe substructure in the residuals to the best-fitting linear phase curve function, which extends beyond the associated uncertainties of the brightness measurements, potentially due to rotational modulation. 

{To search for periodicities in the lightcurve, we use the Lomb-Scargle periodogram algorithm \citep[]{1976Ap&SS..39..447L,1982ApJ...263..835S} 
due to the irregular time spacing of the ATLAS observations.}
For each filter, the best fit linear phase curve is subtracted from each dataset. The resulting data, corrected also for light travel time, were then passed into the Lomb Scargle periodogram function (\emph{LombScargle}) from the \emph{astropy Python} package \citep[]{2013A&A...558A..33A,2018AJ....156..123A}, weighted by the reduced magnitude uncertainty. Frequency grids were chosen with minimum and maximum rotation frequencies of 0.1 day$^{-1}$ (240 hr period) and 24 day$^{-1}$ (1 hr period), with $10^7$ periodogram evaluations between these frequency extremes. Potential rotation periods for each object in the sample were searched for in the $c$ and $o$  datasets separately. For a given filter, 
the maximum peak in the power spectrum was taken to be the object's best fitting period value, with its associated uncertainty equal to half the difference between the period values corresponding to the closest two stationary points either side the maximum peak.

We highlight that the best-fit period value is subject to uncertainty due to aliasing, background noise and uncertainties in the measured magnitudes \citep{2018ApJS..236...16V}.
A method of quantifying the validity of a given rotation period is to calculate the false alarm probability of the maximum peak of the power spectrum - the probability that Gaussian noise could produce a signal of equal height to a given peak in the power spectrum  \citep{2018ApJS..236...16V}. 
For the rotation period value yielded for each filter, we calculated false alarm probabilities,
choosing the Baluev method \citep{2008MNRAS.385.1279B} for its suitable accuracy and computational efficiency \citep{2018ApJS..236...16V}. Only objects whose periods in both filters came from peaks higher than the 95\% false alarm probability level (${>}2\sigma$ results) were considered to be plausible.

Large photometric uncertainties in the data, plus the generally fewer $c$ filter {observations} compared to the $o$-filter, could mean that significantly different best-fit period values are yielded per filter. We set a criterion that the period values from both filters must match to ${<}0.5$ hours to be considered valid.
For any object which fits this criterion, we check if their rotation periods match values aliased to Earth's sidereal period (12, 24, 48, etc. hours) to ${<}0.1$ hours, to ensure we are not detecting periodicity due to sampling rate. If so, we check the rotation phase coverage of the data when phase-folded to the rotation period in each filter. 
If the phase-folded data of an object are grouped into clusters with gaps in between, or cover ${<}50$\% of the rotation phase span, we reject that object's rotation period value.

Three objects in our sample satisfy the above criteria for valid rotation periods: Pluto, Haumea, and Hidalgo. For these objects, we hence consider only the rotation period yielded by the filter with more datapoints ($o$). The ATLAS rotation period for Pluto is consistent {with values from literature} \citep{1994Icar..108..200T,1997Icar..125..245T,2002AJ....124.1757S,2008Icar..197..590S,2010AJ....139.1117B,2018Icar..314..265B} and, to within uncertainties, the ATLAS rotation periods equal the half-values of the literature rotation periods of Haumea \citep[]{2006ApJ...639.1238R,2008AJ....135.1749L,2010A&A...518L.147L,2010A&A...522A..93T,2014EM&P..111..127L} and Hidalgo \citep[]{2016AA...586A.108H,2017AA...604A..95S}.
We note that uncertainties in our reduced magnitude measurements could obscure double-peaked {lightcurves} of elongated or irregular-shaped objects, causing them to appear as single-peaked and the rotation period value yielded by the Lomb-Scargle algorithm to equate to half the true rotation period. Both Haumea \citep[]{2006ApJ...639.1238R,2008AJ....135.1749L,2010A&A...518L.147L,2010A&A...522A..93T,2014EM&P..111..127L} and Hidalgo \citep[]{2007A&A...465..331D,2016AA...586A.108H} exhibit shape elongation. Our measured period values for these objects are consistent with the half the reported rotation periods, thus we consider the true rotation period of these two objects to be twice the value output from the Lomb-Scargle Periodogram. The ATLAS rotation period values for Pluto, Haumea and Hidalgo, both those yielded by the maximum periodogram peak and the final chosen rotation period values, are listed in Table \ref{tablerotation}, in addition to those from literature. 

\begin{figure}
\centering
\includegraphics[width=\textwidth]{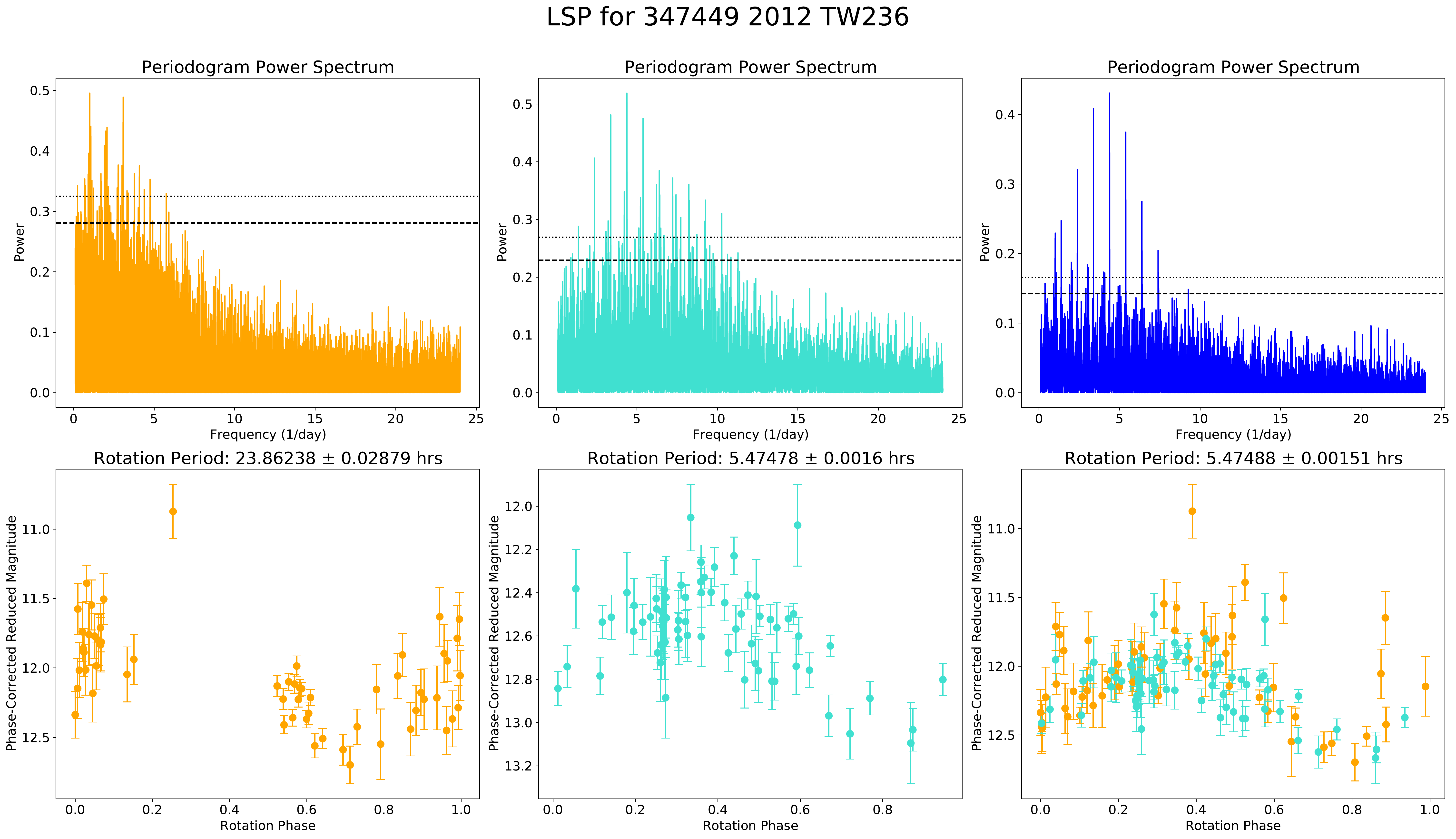}
\caption{{\emph{Upper row:} Lomb-Scargle periodogram power spectra for 2012 TW236 in the $o$ (orange), $c$ (cyan), and combined $c$ and $o$ (blue) filters. The black dashed and dotted lines represent the 95\% and 99.7\% false alarm probability levels. \emph{Lower row:} The $o$ (orange, left), $c$ (cyan, centre), and combined $c$ and $o$ (right) datasets of 2012 TW236, each phase-folded to the rotation period value corresponding to maximum peak in the power spectrum. The combined filter dataset shows a clear sinusoidal lightcurve.}}
\label{LC2012TW236}
\end{figure}

{Some rotation periods may not appear in individual filter datasets, but may be discerned when the datasets are combined to augment sampling of any rotational lightcurve. We repeat our Lomb-Scargle periodogram algorithm on both the ATLAS $c$ and $o$ filter data combined. We assume that the objects in our sample exhibit no variability in colour over their rotation periods, and correct the dataset in each filter for color according to the difference in the median reduced magnitudes of each filter per object. We consider only Lomb-Scargle periodogram spectra maxima higher than the 95\% false alarm probability level. For objects which satisfy this criterion, we consider only those where the data in both filters span the full rotation phase space (ensuring that the rotation period value is not being driven by one filter only) and where the standard deviation of the datapoints exceeds twice the median magnitude uncertainty. If the object's phase-folded lightcurve exhibits gaps in the phase-folded lightcurve, and/or the rotation period value lies close to those aliased to Earth's sidereal period (12, 24, 48 etc. hours) to ${<}0.1$ hours, we reject the object's rotation period value.} {We recover rotation periods for Pluto, Haumea, and Hidalgo, all consistent to those found by separate analysis of each filter. We discover a rotation period value of $5.4749 \pm 0.0015$ hours from the Lomb-Scargle periodogram for the JFC 2012 TW236. The Lomb-Scargle periodogram power spectra and the datasets of 2012 TW236 phase-folded to the obtained rotation periods are shown in Figure \ref{LC2012TW236}. Due to the small size of this object, it is likely irregularly shaped with a double-peaked rotational lightcurve, so we consider the rotation period of 2012 TW236 to be $10.950 \pm 0.003$ hours, to our knowledge the first measured value for this object.}

\begin{figure}
\centering
\includegraphics[width=\textwidth]{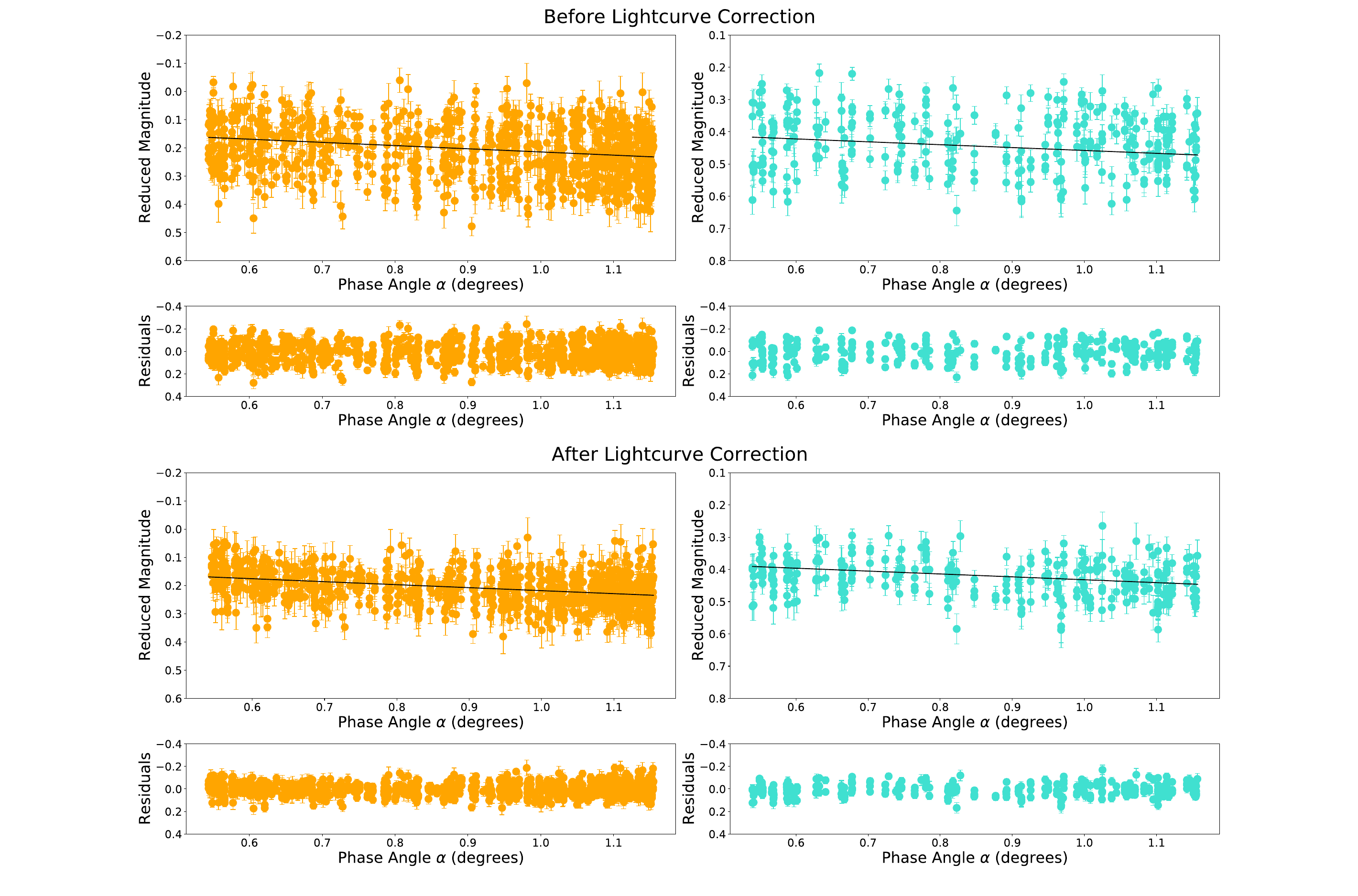}
\caption{\emph{c} and \emph{o} phase curves of Haumea, before (upper plots) and after (lower plots) correction for rotational modulation. Linear best-fit functions plotted as black line, with residuals from the fit displayed beneath each plot.}
\label{LCCorr}
\end{figure}

\begin{deluxetable*}{lcccr}
\tablecaption{Changes in phase coefficient and absolute magnitude in each ATLAS filter for objects whose datasets were corrected for rotational lightcurve. \label{betachange}}
\tablehead{
\colhead{Name} & \colhead{$\beta_{c}$ before} & \colhead{$\beta_{c}$ after} & \colhead{$\beta_{o}$ before} & \colhead{$\beta_{o}$ after} \\
\colhead{} & \colhead{lightcurve correction} & \colhead{lightcurve correction} & \colhead{lightcurve correction} & \colhead{lightcurve correction}\\
\colhead{} & \colhead{(mag/deg)} & \colhead{(mag/deg)} & \colhead{(mag/deg)} & \colhead{(mag/deg)} \\
}
{
\startdata
(134340) Pluto&$0.045\pm0.001$&$0.047\pm0.001$&$0.0460\pm0.006$&$0.0444\pm0.006$\\
(136108) Haumea&$0.056\pm0.016$&$0.090\pm0.016$&$0.113\pm0.010$&$0.106\pm0.011$\\
(944) Hidalgo&$0.0324\pm0.0003$&$0.0318\pm0.0003$&$0.0301\pm0.0002$&$0.0316\pm0.0001$\\
(347449) 2012 TW236&$0.042\pm0.005$&$0.049\pm0.005$&$0.038\pm0.008$&$0.048\pm0.008$\\
\hline
\colhead{Name} & \colhead{$H_{c}$ before} & \colhead{$H_{c}$ after} & \colhead{$H_{o}$ before} & \colhead{$H_{o}$ after}\\
\colhead{} & \colhead{lightcurve correction} & \colhead{lightcurve correction} & \colhead{lightcurve correction} & \colhead{lightcurve correction}\\
\colhead{} & \colhead{(mag/deg)} & \colhead{(mag/deg)} & \colhead{(mag/deg)} & \colhead{(mag/deg)}\\
\hline
(134340) Pluto&$-0.7305\pm0.0015$&$-0.7313\pm0.0014$&$-1.1542\pm0.0007$&$-1.1534\pm0.0007$\\
(136108) Haumea&$0.368\pm0.014$&$0.343\pm0.015$&$0.1011\pm0.0097$&$0.1124\pm0.0097$\\
(944) Hidalgo&$11.051\pm0.009$&$11.074\pm0.009$&$10.776\pm0.004$&$10.744\pm0.004$\\
(347449) 2012T W236&$12.549\pm0.041$&$12.437\pm0.042$&$12.18\pm0.05$&$12.12\pm0.05$\\
\enddata
}
\end{deluxetable*}

\begin{deluxetable*}{lcccccccr}
\tablecaption{Rotation Period obtained from ATLAS datasets in $o$ filter and combined filters. \label{tablerotation}}
\tablehead{
\colhead{Name} & \colhead{$P$} & \colhead{$P_{final}$} & \colhead{Filter(s)} & \colhead{$\Delta M$ $c$} & \colhead{$\Delta M$ $o$} & \colhead{Literature $P$} & \colhead{Literature $\Delta M$} & \colhead{Source}\\
\colhead{} & \colhead{(hr)} & \colhead{(hr)} & \colhead{} & \colhead{(mag)} & \colhead{(mag)} & \colhead{(hr)} & \colhead{(mag)} & \colhead{}
}
\startdata
(134340) Pluto & $153.28\pm0.65$ & $153.28\pm0.65$ & $o$ & $0.100\pm0.004$ & $0.080\pm0.002$ & $153.2328\pm0.0048$ & $0.120\pm0.006$ & B18\\
(136108) Haumea & $1.95767\pm0.0001$ & $3.91534\pm0.00019$ & $o$ & $0.19\pm0.01$ & $0.205\pm0.004$ & $3.915341\pm0.000005$ & $0.28\pm0.02$ & SS17\\
   (944) Hidalgo & $5.031\pm0.004$ & $10.06\pm0.009$ & $o$ & $0.101\pm0.031$ & $0.237\pm0.014$ & $10.0630\pm0.0003$ & $0.48$ & W97 \\
{(347449) 2012 TW236} & {$5.4749\pm0.0015$} & {$10.950\pm0.003$} & {$c$ and $o$} & {$0.444\pm0.072$} & {$0.568\pm0.116$} &{...}&{...}&{...}\\
& & & {combined} & & & & &\\
\enddata
\tablerefs{Rotational period/amplitude sources: 
B18: \citet{2018Icar..314..265B}, 
SS17: \citet{2017AA...604A..95S}, 
W97: \citet{1997Icar..126..395W}.}
\end{deluxetable*}

We phase-fold the datasets of Pluto, Haumea, Hidalgo{, and 2012 TW236} in both filters to their selected rotation period value. 
We fit a sinusoid of period equal to the chosen rotation period value to the data using the
\emph{lmfit} Python package \citep{2014zndo.....11813N}, weighted by the uncertainty of each magnitude. 
This fitted function is subtracted off the reduced magnitude data to eliminate the rotational modulation, and {remove any outlying data in the corrected dataset to 3-sigma. T}he linear phase curve model for the object is refitted again on the corrected data to measure the phase coefficient and absolute magnitude, corrected for the effect of the object's rotation on its axis. The effect of correcting for an object's rotational {lightcurve} is illustrated in Figure \ref{LCCorr}, showing the significant decrease in spread of the data about the best fitting linear phase curve function. The root mean square values of the residuals to the best-fitting linear function decrease from 0.096 mag to 0.066 mag in the $c$ filter and from 0.093 mag to 0.057 mag in the $o$ filter when the effects of Haumea's rotational {lightcurve} is removed. 
{Table \ref{betachange} lists the phase coefficient and absolute magnitude values for each object for which rotational modulation was corrected, before and after correction.}
Table \ref{tablerotation} lists the amplitudes of the best-fitting sinusoid {lightcurve} function in both $c$ and $o$, the uncertainty being calculated from the covariance matrix of the resulting fit.
We note that our amplitudes of rotational modulation differ significantly both between filters and from those in literature, though the large uncertainties in ATLAS reduced magnitudes are a possible cause of this.

\subsection{Cometary Activity Search} \label{CometActivityCorr}

The Centaur and JFC populations contain objects known to exhibit cometary activity {\citep[and references therein]{2008ssbn.book..397L,2020tnss.book..307P}}. It is thus plausible that some objects in our sample may have exhibited activity, forming a coma which may partly or fully obscure an object's surface from direct observation. This would brighten the object's apparent magnitude, and could alter an object's phase curve for a period of time. Given the number of datapoints per object, we can use the ATLAS phase curves to search for deviations that may be indicative of possible cometary outgassing or dust production. To search for cometary outbursts, we {select} for observations that were significantly brighter than predicted from the measured phase curve. We assume that most of an object's data are taken from epochs of (relative) quiescence potentially punctuated by shorter epochs of activity or outbursts.
We consider this assumption to be valid as most of our objects exhibit reduced magnitudes which are consistent in value across separate apparitions.
We subtract the best-fitting phase curve function from the dataset (corrected for rotational {lightcurve} if possible), and select observations brighter than $2\sigma$ from the object's median brightness across the ATLAS baseline. 

Active epochs of Centaurs and JFCs exhibit timescales ranging from weeks to years {\citep[and references therein]{2020tnss.book..307P}}.
Such activity would manifest as a series of bright observations across consecutive nights. 
To detect such cometary activity, we utilize the ability of ATLAS to observe a given target multiple times per night, and select nights containing at least 3 observations, of which ${\geq}75\%$ have magnitudes brighter than $2\sigma$ from the object's median brightness. We term such nights of observations `bright nights' henceforth. We further analyse these bright nights to determine if cometary activity is the source of the brightening. We note that objects may exhibit activity which do not produce short increases in brightness, and we would not be able to detect this activity using the described algorithm. To detect any such lower-level cometary activity, we consider only observations whose phase-corrected reduced magnitude lie ${<}2\sigma$ from the object's sigma-clipped median brightness. We consider each datapoint in turn, and bin all data taken 30 days after each observation. If at least $14$ observations (the approximate average number of ATLAS exposures per week) are taken in any given interval, we consider the median brightness of all data in that interval. If the binned median brightness exceeds the object's sigma-clipped median brightness by $0.2$ mag and $1$ standard deviation of the binned data, we consider all data in this interval to be a long outburst.
We remove any data which corresponded to epochs of cometary activity from the datasets and reapply the phase curve fitting algorithm to the remaining data. {If any instance of cometary activity is found to extend over time, we remove all data in both filters that falls within the time range of the activity.}

\subsection{HG$_{1}$G$_{2}$ Model} 

Several objects in our sample were observed across phase angle ranges comparable to those of main belt asteroids. This allowed us to fit their data from ATLAS with phase curve functions used for these asteroids. The resulting best-fit parameters for our sample could then be compared to those of the asteroid population. We utilize the three-parameter $HG_{1}G_{2}$ phase curve model developed by \citet{2010Icar..209..542M}, where $H$ is the absolute magnitude of the phase curve, and $G_1$ and $G_2$ both quantify the slopes of the phase angle functions that govern the overall phase curve shape, analogous to the linear phase coefficient $\beta$. {We utilize this function as per the findings of \citet{2021Icar..35414094M} that taxonomic complexes could be distinguished using the $HG_{1}G_{2}$ function applied to ATLAS data. \citet{2021Icar..35414094M} also found the related $HG_{12}^{*}$ asteroid phase curve function \citep{2016P&SS..123..117P} failed to distinguish between complexes for ATLAS data, thus we do not apply the $HG_{12}^{*}$ function to our data.} We select only phase curves in a given filter to fit with the $HG_{1}G_{2}$ model which satisfy the following criteria:

\begin{enumerate}
    \item Observations of $\alpha \geq 12$ deg, to ensure we are sampling the section of the phase curve where curvature becomes significant and the accuracy of linear fits is reduced.
    \item Phase curves must contain ${\geq}35$ datapoints at phase angles $0 \leq \alpha \leq 5$ deg, to ensure we have sufficient sampling of the region where the opposition surge can become significant, which may affect the profile of the fitted phase curve model.
\end{enumerate}

The $o$ filter phase curves of Narcissus, Hidalgo, and 2014 QA43, and the $c$ filter phase curve of Hidalgo satisfy these criteria. We apply the $HG_{1}G_{2}$ model to {the reduced magnitude values} of these datasets using the \emph{HG1G2} function from the \emph{Python} photometry module \emph{sbpy.photometry} \citep{2019JOSS....4.1426M}{. We use} a Monte Carlo algorithm 
to measure the absolute magnitude and $G_1$ and $G_2$ parameters, generating $10^5$ synthetic phase curves by varying each reduced magnitude value by a random fraction of its associated uncertainty (assuming a {normal} uncertainty distribution) and fitting the $HG_{1}G_{2}$ model to each synthetic dataset. The final values for $H$, $G_{1}$, and $G_{2}$ are the median of the resulting distribution of all $10^5$ values of each parameter, with their associated uncertainties being half of the range between the values which enclose 95\% of all values in the distribution, yielding $2\sigma$ {uncertainties}. If an object's data satisfy the requirements for fitting with the $HG_{1}G_{2}$, we retroactively use the resulting fit instead of the linear model for eliminating phase angle dependence when searching for rotation periods and cometary activity in the full dataset. 
For the object Narcissus, due to near-identical resulting chi-square values, we use both the linear and $HG_{1}G_{2}$ fits when searching for rotation period and cometary activity, finding that the choice of fitted function has no effect on the results yielded for Narcissus by either algorithm.

\section{Results and Discussion} 

\subsection{Linear Phase Coefficients - Full Phase Angle Range}

We measured the linear phase coefficient $\beta$ and absolute magnitude $H$ in both ATLAS $c$ and $o$ filters for all 18 objects in our sample {across the full range of phase angles spanned by each dataset (excluding opposition surges at $\alpha{<}0.1$ deg).} 
Table \ref{linearbetatable} lists our measured 
$\beta$ and 
$H$ values and $2\sigma$ errors for all objects in our sample {across the full phase angle range}. 
Figures for all $c$ and $o$ ATLAS phase curves {and the
2-dimensional (2D) histograms of the fitted phase curve
parameters} are shown in Figure \ref{PhCExampleQuaoar} and its online Figure set.
{Linear fits applied to the full range of phase angles spanned by the data were used to correct for variability due to object rotation and cometary activity for all objects, apart from the $o$ filter datasets of Narcissus and 2014 QA43 and the $c$ and $o$ filter datasets of Hidalgo which were corrected using the $HG_{1}G_{2}$ function. 
Four such objects - 2012 TW236, 2014 QA43, 2016 ND21, and Narcissus - have datasets which extend to phase angles beyond $\alpha\sim10$ deg where non-linearity could become significant, but 
the small number of objects this applies to means we consider a linear fit suitable for variability correction. However, for the remainder of our analysis, including a 
comparison to the Kuiper belt and is associated populations, we restrict ourselves to measuring phase coefficients and absolute magnitudes for a fixed phase angle range of $0.1\leq\alpha\leq6$ deg, as detailed in section \ref{LinearBeta0_6_deg}.}


\begin{longrotatetable}
\begin{deluxetable*}{lcccccccccr}
\tablecaption{Linear phase coefficients $\beta$ and absolute magnitudes $H$ for all objects in our sample in both ATLAS broadband filters \emph{c} and \emph{o}, measured over the full range of phase angles for each filter dataset, with associated $2\sigma$ uncertainties, with phase angle range over which the parameters were measured in each filter. Parameter values measured across the maximum range of phase angles in each filter, with data corrected for rotational lightcurves and cometary activity wherever possible.\label{linearbetatable}}
\tablehead{
\colhead{Object} & \colhead{$\beta_{c}$} & \colhead{$H_{c}$} & \colhead{$\alpha$ range ($c$ filter)} & \colhead{$N_{c}$} & \colhead{$\beta_{o}$} & \colhead{$H_{o}$} & \colhead{$\alpha$ range ($o$ filter)} & \colhead{$N_{o}$} &  \colhead{Light} & \colhead{Cometary}\\
\colhead{} & \colhead{(mag/deg)} & \colhead{(mag)} & \colhead{(deg)} & \colhead{} & \colhead{(mag/deg)} & \colhead{(mag)} & \colhead{(deg)} & \colhead{} & \colhead{curve} & \colhead{activity}\\
\colhead{} & \colhead{} & \colhead{} & \colhead{} & \colhead{} & \colhead{} & \colhead{} & \colhead{} & \colhead{} & \colhead{correction?} & \colhead{correction?}
}
{
\startdata
\multicolumn{11}{c}{KBOs}\\
\hline
(136199) Eris&$0.043\pm0.088$&$-1.012\pm0.042$&0.1206-0.6058&239&$0.016\pm0.043$&$-1.307\pm0.018$&0.1214-0.6068&738&No&No\\
(136108) Haumea&$0.089\pm0.017$&$0.343\pm0.014$&0.5397-1.1567&313&$0.106\pm0.01$&$0.112\pm0.01$&0.544-1.1548&1043&Yes&No\\
(38628) Huya&$-0.015\pm0.085$&$4.974\pm0.096$&0.5515-1.9664&68&$0.058\pm0.031$&$4.476\pm0.04$&0.5394-2.0139&340&No&No\\
(136472) Makemake&$0.005\pm0.015$&$0.107\pm0.013$&0.5209-1.1062&353&$0.034\pm0.009$&$-0.247\pm0.008$&0.5193-1.1087&1088&No&No\\
(90482) Orcus&$0.18\pm0.065$&$2.238\pm0.059$&0.4008-1.2082&271&$0.07\pm0.04$&$2.07\pm0.037$&0.4045-1.2084&541&No&No\\
(134340) Pluto{$^a$}&$0.045\pm0.001$&$-0.728\pm0.002$&0.1085-1.7313&323&$0.0448\pm0.0006$&$-1.1536\pm0.0007$&0.1152-1.7362&886&Yes&No\\
(50000) Quaoar&$0.029\pm0.065$&$2.792\pm0.057$&0.1964-1.341&177&$0.075\pm0.028$&$2.221\pm0.025$&0.1925-1.3402&590&No&No\\
\hline
\multicolumn{11}{c}{Centaurs}\\
\hline
(459865) 2013 XZ8&$0.026\pm0.017$&$9.937\pm0.087$&2.8174-6.8551&127&$0.036\pm0.011$&$9.524\pm0.052$&2.1459-6.9036&329&No&No\\
(54598) Bienor&$0.058\pm0.025$&$7.586\pm0.07$&1.3222-4.0222&172&$0.062\pm0.014$&$7.248\pm0.041$&1.3103-4.0452&431&No&No\\
(10199) Chariklo&$0.035\pm0.019$&$6.939\pm0.044$&0.1041-3.6314&170&$0.049\pm0.01$&$6.477\pm0.022$&0.1535-3.5772&454&No&No\\
(2060) 95P/Chiron&$0.093\pm0.011$&$5.69\pm0.023$&0.2194-3.093&234&$0.097\pm0.006$&$5.433\pm0.011$&0.1688-3.1055&834&No&Yes\\
\hline
\multicolumn{11}{c}{JFCs}\\
\hline
(347449) 2012 TW236&$0.049\pm0.005$&$12.437\pm0.042$&2.0318-17.0886&69&$0.048\pm0.008$&$12.123\pm0.05$&1.9453-14.469&58&Yes&No\\
(501585) 2014 QA43&$0.063\pm0.015$&$11.781\pm0.131$&3.7953-11.9905&81&$0.064\pm0.005$&$11.379\pm0.041$&1.2204-11.8886&270&No&Yes{$^b$}\\
2016 ND21&$0.032\pm0.005$&$12.635\pm0.043$&0.1533-14.7265&91&$0.048\pm0.003$&$12.063\pm0.027$&2.0524-15.0726&333&No&No\\
(944) Hidalgo&$0.0318\pm0.0003$&$11.074\pm0.009$&2.1269-30.5594&178&$0.0316\pm0.0002$&$10.744\pm0.004$&1.1296-30.5823&730&Yes&No\\
(37117) Narcissus&$0.048\pm0.003$&$13.573\pm0.023$&2.5505-18.5841&69&$0.044\pm0.002$&$13.161\pm0.021$&1.5666-18.3289&204&No&No\\
\hline
\multicolumn{11}{c}{Transition Objects}\\
\hline
(349933) 2009 YF7&$0.032\pm0.018$&$11.117\pm0.099$&1.9905-8.5739&110&$0.036\pm0.01$&$10.747\pm0.064$&1.9827-8.9607&285&No&No\\
(60558) 174P/Echeclus&$0.083\pm0.008$&$9.804\pm0.039$&0.9073-8.6663&111&$0.064\pm0.003$&$9.403\pm0.015$&0.383-9.5528&363&No&Yes\\
\enddata
}
\tablenotetext{^a}{Values are for the Pluto/Charon system, whose components cannot be resolved at the ATLAS pixel scale.}
\tablenotetext{^b}{While we do not detect any confirmed instance of cometary activity for 2014 QA43, we nevertheless remove all data classified by our activity search algorithm as occurring during an instance of sustained brightening. It is not clear if this brightening is due to cometary activity not visible in the ATLAS images or due to changing aspect angle.}
\end{deluxetable*}
\end{longrotatetable}

\subsection{Linear Phase Coefficients - Limited Phase Angle Range}\label{LinearBeta0_6_deg}

\begin{figure}
\centering
\includegraphics[width=\columnwidth]{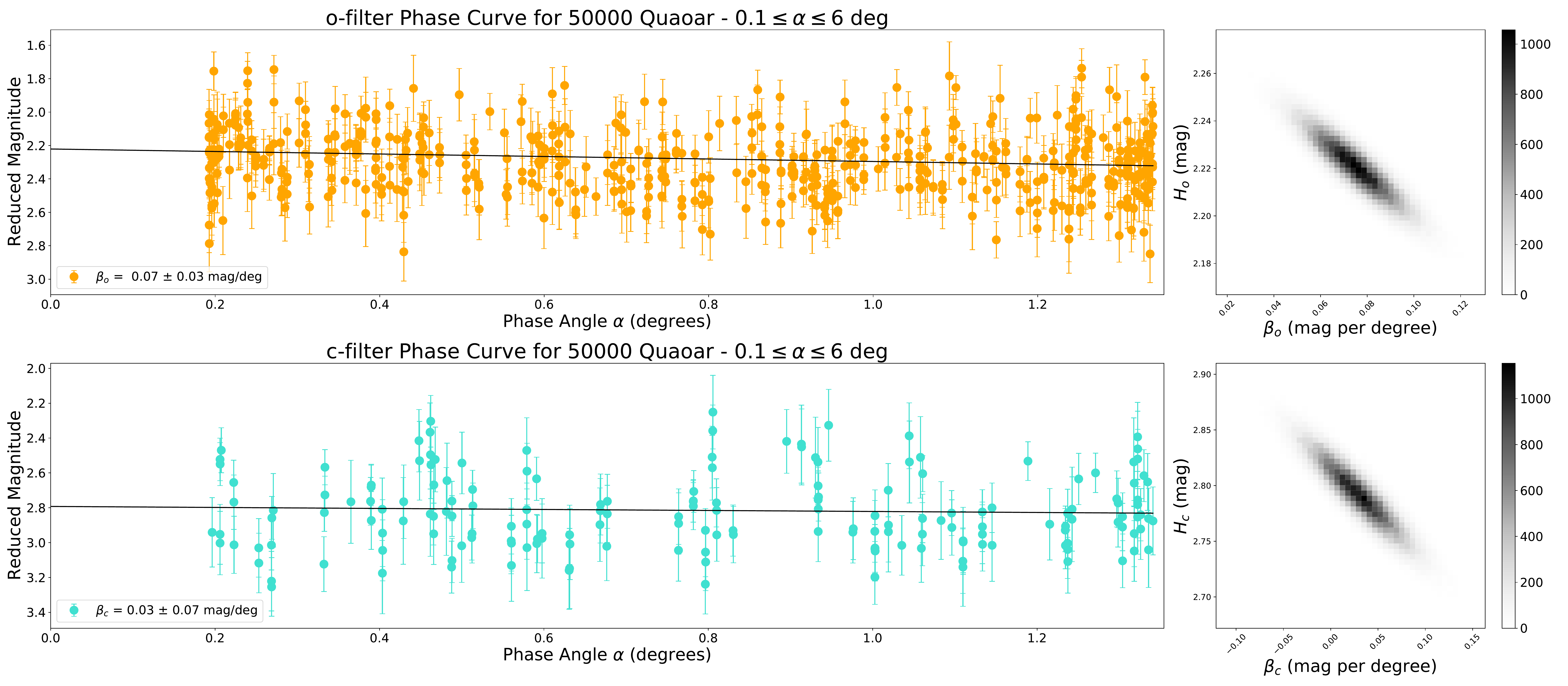}
\caption{{Linear fit to phase curve of (50000) Quaoar in ATLAS
    $o$ and $c$ filters for data in phase angle range
    $0.1\leq\alpha\leq6$ deg. Best fit function extended to $\alpha =
    0$ deg for ease of viewing the extrapolated absolute magnitude
    value and any potential opposition surge.}
    The complete figure set (13 images) is available in the online published version of the paper.}
\label{PhCLinearQuaoar_0_6_deg}
\end{figure}

The objects in our sample span semimajor axes of $5.7$ au $\leq a \leq$ $68.1$ au, resulting in a wide range of maximum phase angles observable from Earth. ATLAS observations thus sample different sections of the phase curves depending on an object's heliocentric distance. To compare phase curves over a fixed range of phase angle, we restrict our datasets (corrected for rotational modulation and cometary activity where possible and excluding data at $\alpha\leq0.1$ deg) to include only observations taken at phase angles $\alpha\leq{6}$ deg, and consider only objects with ${>}50$ datapoints in this range in both $c$ and $o$. We apply our methods to the resulting phase curves of {13} objects. The resulting absolute magnitudes and phase coefficients in $c$ and $o$ are listed in Table \ref{linearbetatable6deg}. 
{Comparing the resulting values to those measured over the maximum
phase angle range, w}e find that the phase coefficient and
absolute magnitude values change beyond $2\sigma$ error bars
{only for the Transition Objects Echeclus and 2009 YF7 in the $c$
filter. Figures for all $c$ and $o$ ATLAS phase curves for data
in phase angle range $0.1\leq\alpha\leq{6}$ are shown in Figure
\ref{PhCLinearQuaoar_0_6_deg} and its online Figure set.} 

\begin{deluxetable*}{lcccccccr}
\tablecaption{Linear phase coefficients $\beta$ and absolute magnitudes $H$ for all objects with ${>}50$ data points recorded at phase angles $0.1\leq\alpha\leq6$ deg in both ATLAS broadband filters \emph{c} and \emph{o}, with associated $2\sigma$ uncertainties, and number of data points $N$ at $0.1\leq\alpha\leq6$ deg in each ATLAS filter. \label{linearbetatable6deg}}
\tablehead{
\colhead{Object} & \colhead{$\beta_{c}$} & \colhead{$H_{c}$} & \colhead{$N_{c}$} & \colhead{$\beta_{o}$} & \colhead{$H_{o}$} & \colhead{$N_{o}$} & \colhead{Light} & \colhead{Cometary}\\
\colhead{} & \colhead{(mag/deg)} & \colhead{} & \colhead{($\alpha\leq5$ deg)} & \colhead{(mag/deg)} & \colhead{} & \colhead{($\alpha\leq5$ deg)} & \colhead{curve} & \colhead{activity}\\
\colhead{} & \colhead{} & \colhead{} & \colhead{} & \colhead{} & \colhead{} & \colhead{} & \colhead{correction?} & \colhead{correction?}
}
{
\startdata
\multicolumn{9}{c}{KBOs}\\
\hline
(136199) Eris& $0.043 \pm 0.088$  & $-1.012 \pm 0.042$ & 239 & $0.016 \pm 0.043$ & $-1.307 \pm 0.018$ & 738  & No& No\\
(136108) Haumea& $0.089 \pm 0.017$  & $0.343 \pm 0.014$  & 313 & $0.106 \pm 0.01$  & $0.112 \pm 0.01$   & 1043 & Yes& No\\
(38628) Huya& $-0.015 \pm 0.085$ & $4.974 \pm 0.096$  & 68  & $0.058 \pm 0.031$ & $4.476 \pm 0.040$  & 340  & No& No\\
(136472) Makemake& $0.005 \pm 0.015$  & $0.107 \pm 0.013$  & 353 & $0.034 \pm 0.009$ & $-0.247 \pm 0.008$ & 1088 & No& No\\
(90482) Orcus& $0.18 \pm 0.065$   & $2.238 \pm 0.059$  & 271 & $0.07 \pm 0.04$   & $2.07 \pm 0.037$   & 541  & No& No\\
(134340) Pluto{$^a$}& $0.045 \pm 0.001$  & $-0.728 \pm 0.002$ & 323 & $0.0448 \pm 0.0006$ & $-1.1536 \pm 0.0007$ & 886  & Yes& No\\
(50000) Quaoar& $0.029 \pm 0.065$  & $2.792 \pm 0.057$  & 177 & $0.075 \pm 0.028$ & $2.221 \pm 0.025$  & 590  & No& No\\
\hline
\multicolumn{9}{c}{Centaurs}\\
\hline
(459865) 2013 XZ8& $0.024 \pm 0.023$  & $9.948 \pm 0.105$  & 85  & $0.031 \pm 0.016$ & $9.546 \pm 0.069$  & 236  & No& No\\
(54598) Bienor& $0.058 \pm 0.025$  & $7.586 \pm 0.07$   & 172 & $0.062 \pm 0.014$ & $7.248 \pm 0.041$  & 431  & No& No\\
(10199) Chariklo& $0.035 \pm 0.019$  & $6.939 \pm 0.044$  & 170 & $0.049 \pm 0.01$  & $6.477 \pm 0.022$  & 454  & No& No\\
(2060) 95P/Chiron& $0.093 \pm 0.011$  & $5.69 \pm 0.023$   & 234 & $0.097 \pm 0.006$ & $5.433 \pm 0.011$  & 834  & No& Yes\\
\hline
\multicolumn{9}{c}{Transition Objects}\\
\hline
(60558) Echeclus& $0.094 \pm 0.012$  & $9.769 \pm 0.048$  & 81  & $0.069 \pm 0.006$ & $9.394 \pm 0.018$  & 252  & No& Yes\\
(349933) 2009 YF7& $0.113 \pm 0.033$  & $10.768 \pm 0.159$ & 85  & $0.067 \pm 0.027$ & $10.611 \pm 0.125$ & 137  & No& No\\
\enddata
}
\tablenotetext{^a}{Values are for the Pluto/Charon system, whose components cannot be resolved at the ATLAS pixel scale.}
\end{deluxetable*}

\begin{figure}
\centering
\includegraphics[width=\textwidth]{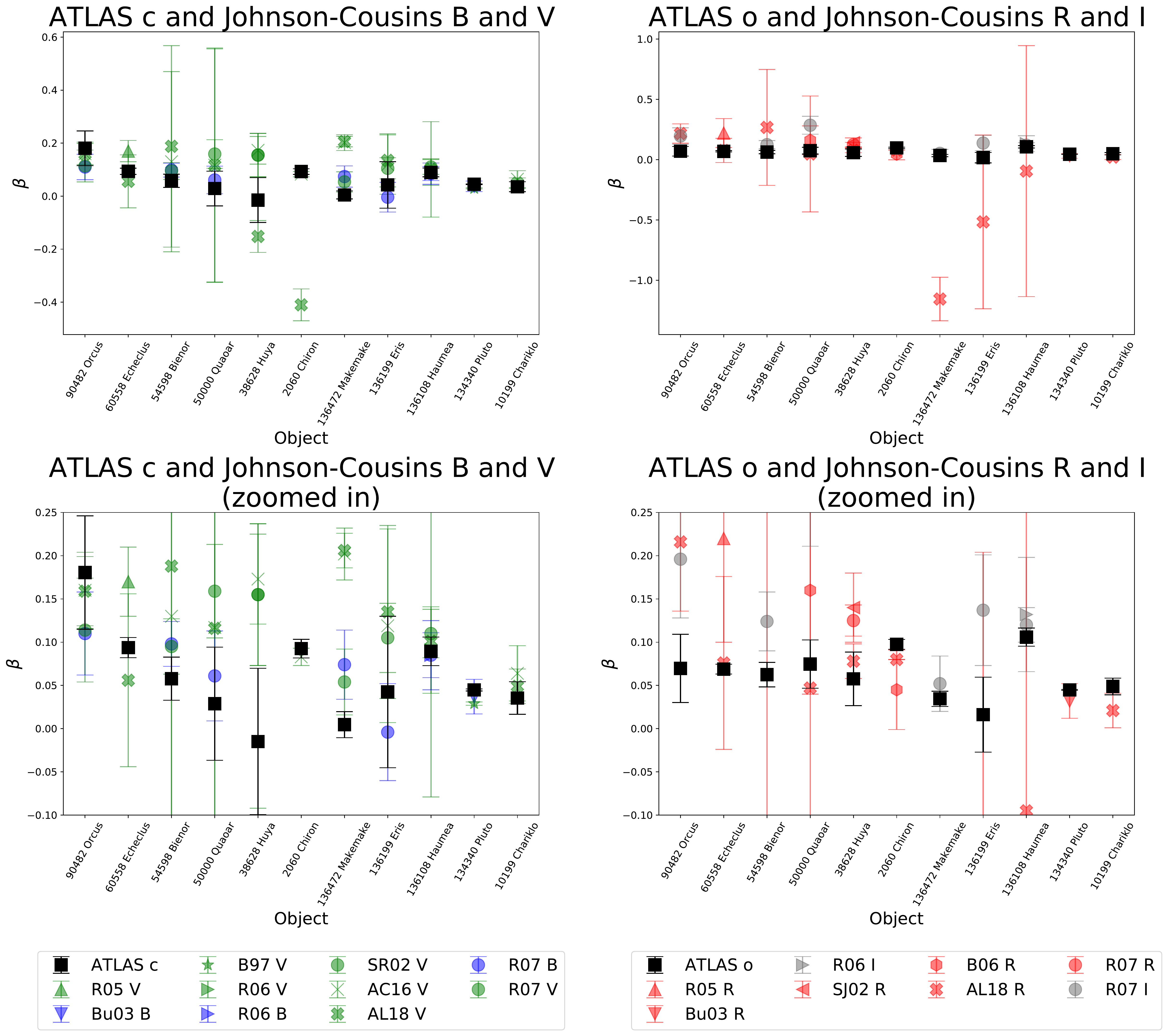}
\caption{\emph{Upper plots:} Linear phase coefficients in ATLAS $c$ (\emph{left}, black squares) and $o$ (\emph{right}, black squares), calculated from the phase angle range $0.1\leq\alpha\leq6$ deg, with literature values in Johnson-Cousins broadband filters (blue: $B$, green: $V$, red: $R$, grey: $I$) with which ATLAS filters exhibit significant overlap. 
\emph{Lower plots:} Upper plots magnified for clarity.
AC16: \citet{2016A&A...586A.155A},
AL18: \citet{2018MNRAS.481.1848A},
Bu03: \citet{2003Icar..162..171B},
B97: \citet{1997Icar..125..233B},
B06: \citet{2006A&A...450.1239B},
R05: \citet{2005Icar..176..478R},
R06: \citet{2006ApJ...639.1238R},
R07: \citet{2007AJ....133...26R},
SJ02: \citet{2002AJ....124.1757S},
SR02: \citet{2002Icar..160...52S}.
\label{BetaLitComparison}}
\end{figure}

We note that our $\beta$ values differ by ${>}2\sigma$ between the ATLAS filters for many objects. Significant variation of $\beta$ with wavelength has been previously reported by several multi-filter phase curve studies of Centaurs and KBOs \citep{2003Icar..162..171B,2007AJ....133...26R,2009AJ....137..129S,2016A&A...586A.155A,2018MNRAS.481.1848A,2019MNRAS.488.3035A}. Figure \ref{BetaLitComparison} shows the ATLAS $\beta$ values compared to those from the literature in the Johnson-Cousins \emph{BVRI} photometric system, for which the ATLAS filters $c$ and $o$ exhibit significant overlap in wavelength range with $B$ and $V$ filters ($c$ filter) and $R$ and $I$ filters ($o$ filter), respectively. Though the ATLAS $\beta$ values often differ in value from the literature, for most objects they overlap within the $2\sigma$ error bars or lie between the extremes of the range of the literature values for a given object. Our ATLAS linear phase coefficients are consistent with those from the literature, as expected, and we attribute any disparity 
to be due to different wavelength ranges of the filters of the ATLAS and Johnson-Cousins photometric systems. We also note that our measured phase coefficient and absolute magnitudes for Pluto are {those} of the Pluto-Charon system, the components of which cannot be resolved at the ATLAS pixel scale.

\begin{figure}
\centering
\includegraphics[width=\columnwidth]{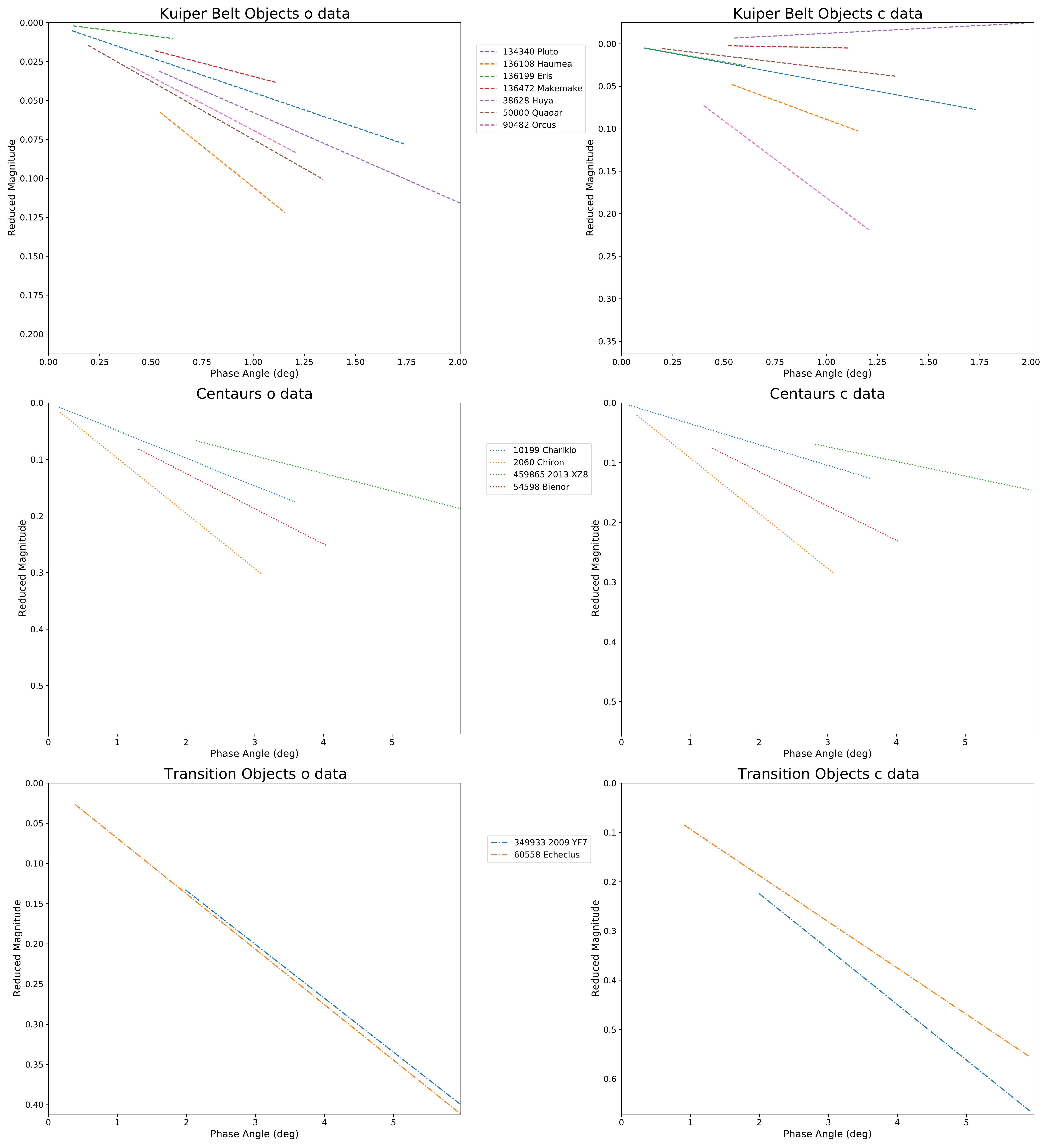}
\caption{Best-fit linear phase curves for all objects in our sample, divided by filter ($c$ filter left, $o$ filter right) and dynamical classification, spanning the phase angle range of each dataset in each filter. All phase curves are normalized to 0 mag at opposition ($\alpha = 0$ deg) to enable comparison of slope values.}
\label{AllPhaseCurves}
\end{figure}

\begin{figure}
\centering
\includegraphics[width=\columnwidth]{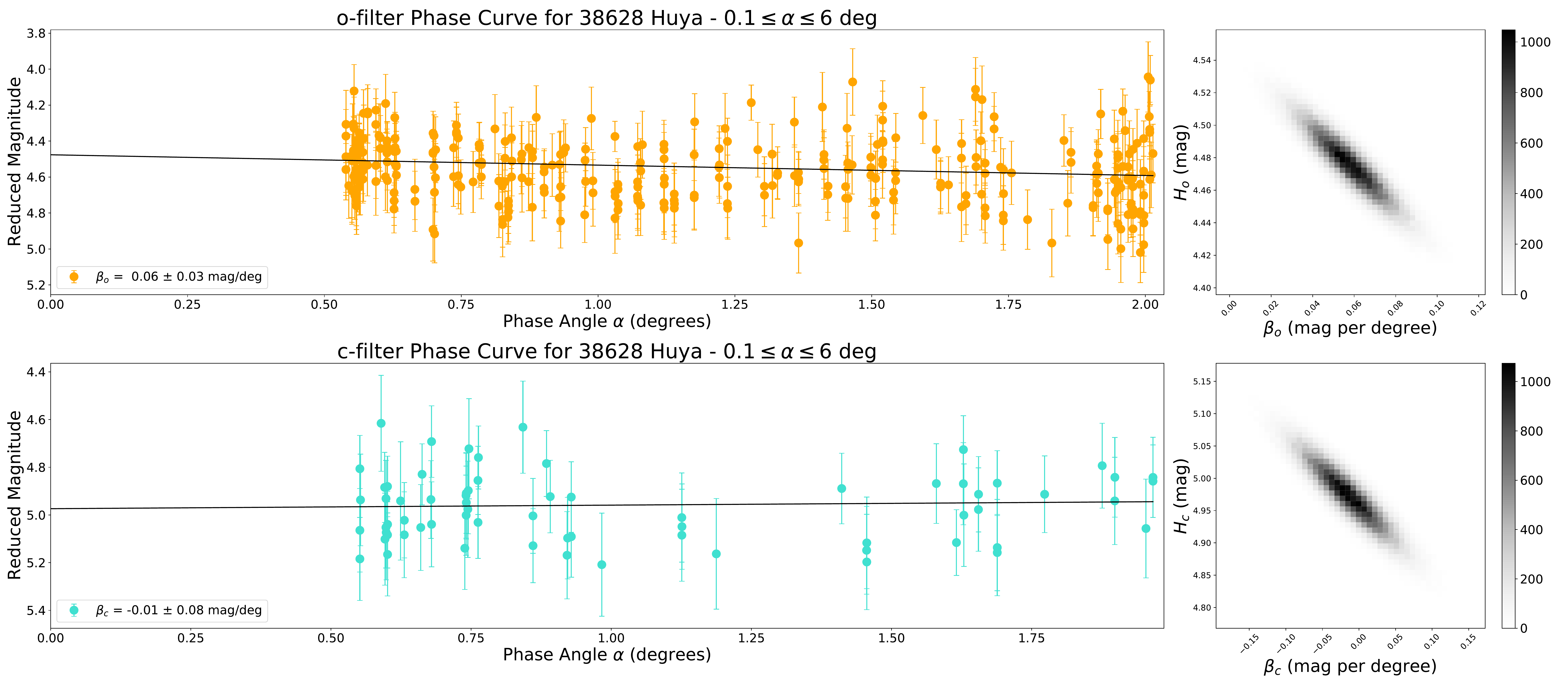}
\caption{Linear fit to phase curve of (38628) Huya in $c$ and $o$ filters, with 2D histograms showing density plot of error distributions of linear phase coefficient and absolute magnitude. Best fit function extended to $\alpha = 0$ deg for ease of viewing any potential opposition surge. Note the nominally negative $\beta_{c}$ value, whose associated uncertainty nevertheless makes it consistent with a flat sloped phase curve, consistent with atmosphereless surface reflectance.}
\label{PhCExampleHuya}
\end{figure}

Figure \ref{AllPhaseCurves} shows the best-fitting linear phase curve models for all objects in our sample, divided by dynamical class and photometric filter. The $\beta$ values indicate relatively flat phase curves across our entire sample, with values ranging from $-0.01$ to $0.18$ mag/deg in $c$, and $0.02$ to $0.106$ mag/deg in $o$. Such flat phase curves have been previously observed for several JFCs, Centaurs, and Pluto-sized KBOs \citep{2006ApJ...639.1238R,2007AJ....133...26R,2017MNRAS.471.2974K,2018MNRAS.479.4665K,2022PSJ.....3...95V} and are thought to be indicative of low-albedo organic-rich Centaur and JFC surfaces \citep{2000Icar..147..545N,2002Icar..159..396S,2003EM&P...92..315C,2007AJ....133...26R} and high-albedo KBO surfaces covered in granular volatile ices \citep[]{2005ApJ...635L..97B,2006A&A...445L..35L,2006ApJ...639.1238R,2007AJ....133...26R,2007ApJ...655.1172T}. {Furthermore, we find that for both the $c$ and $o$ filters, the phase coefficient values of the Centaurs and Transition Objects all fall in the same range as those of the KBO population when measured across the same phase angle range $0.1\leq\alpha\leq6$ deg.}
All but one of our sample exhibit positive phase coefficient slopes in both ATLAS filters. Huya exhibits a marginally negative phase coefficient in the $c$ filter, as shown in Figure \ref{PhCExampleHuya} but is consistent within $2\sigma$ of having a flat or positive slope. \citet[]{2016A&A...586A.155A,2018MNRAS.481.1848A,2019MNRAS.488.3035A} reported negative phase coefficients for several objects in their sample, including Chiron, Eris, Haumea, Huya, and Makemake. $\beta<0$ would imply an unphysical dimming near opposition. \citet{2016A&A...586A.155A} and \citet{2018MNRAS.481.1848A} speculated this effect may be due to unidentified ring systems, epochs of cometary activity, or poorly corrected rotational modulation. We instead measure phase coefficient values for these objects which are either positive slopes or consistent with positive slopes to within $2\sigma$. 
Furthermore, Huya's negative $c$ filter phase coefficient is derived from the second sparsest dataset in our sample of $N_c = 68$ observations, comparable in size to the $V$ filter dataset from \citet{2018MNRAS.481.1848A} ($N_V = 45$) which also yielded a negative phase coefficient value. 
This tendency towards positive $\beta$ values with increasing dataset size suggests negative phase coefficient values are most likely due to small datasets with large scatter which are unable to sufficiently sample the phase curve profile, and not a real property of these objects.

\subsection{Correlations}

\begin{figure}
\centering
\includegraphics[width=0.8\textwidth]{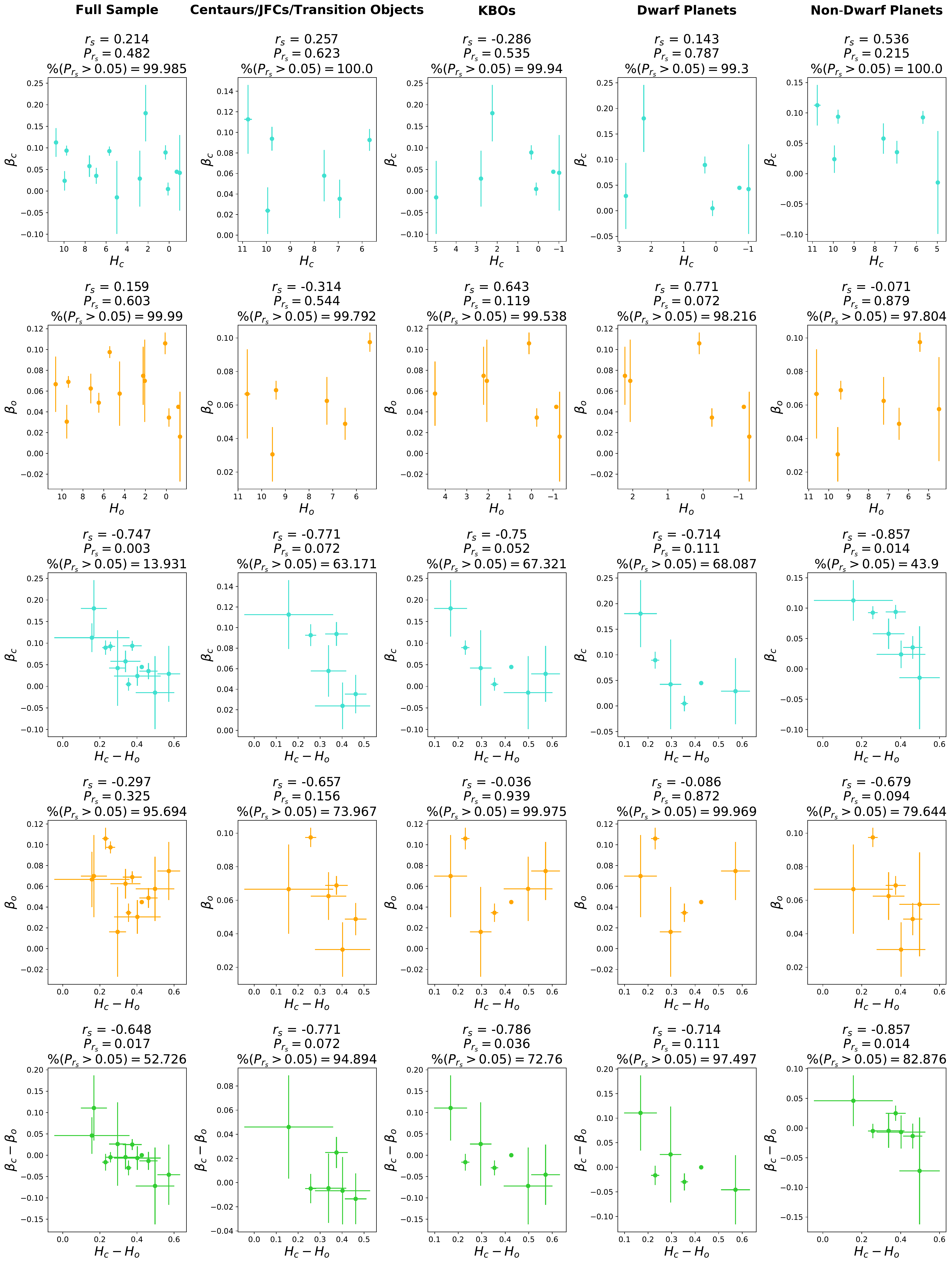}
\caption{Plots of tested parameter pairs derived from ATLAS observations for our full object sample (first column); Centaurs, JFCs, and Transition Objects (second column); KBOs (third column); dwarf planets (fourth column); and non-dwarf planets (last column). 
\label{ATLASCorrelations}}
\end{figure}

We explore potential correlations between phase coefficients $\beta_{c}$ and $\beta_{o}$, absolute magnitudes $H_{c}$ and $H_{o}$, color index $H_{c}-H_{o}$, and relative phase coefficients $\beta_{c}-\beta_{o}$ derived from the ATLAS phase curves. {We analyse parameter values measured across the phase angle range $0.1 \leq \alpha \leq 6$ deg to ensure all parameters are measured over the same region of the phase curve.} We make use of the Spearman rank correlation test to search for monotonic relationships between pairs of variables, using the \emph{spearmanr} function from the python module \emph{scipy} \citep{2020NatMe..17..261V}. The \emph{spearmanr} function calculates two parameters. The first is the Spearman rank correlation coefficient, $r_{s}$, which quantifies the strength and direction of any detected monotonic relationship between two variables; $|r_{s}| \rightarrow 1$ defines a strong correlation, whereas $r_{s} \approx 0$ implies no correlation. The second parameter is the null hypothesis probability value of that correlation, or p-value, $P_{r_{s}}$, which quantifies the probability that any detected correlation is derived from an inherently uncorrelated distribution of data. 
We note however that the Spearman rank correlation test itself does not consider uncertainties in the analysed variables. 
To account for this, for every parameter pair tested, we perform a Monte Carlo simulation on the parameter values. For each data point, we vary the nominal value within associated uncertainties{, assuming every uncertainty follows a normal distribution with a standard deviation equal to the nominal value of the uncertainty itself}, perform the rank statistic on the resulting synthetic values, and repeat $10^5$ times. 
We accumulate all the associated p-values and calculate the percentage of simulations for which $P_{r_{s}} > 0.05$ i.e. we could not rule out an uncorrelated population to 95\% probability. 
We consider a correlation to exist if $|r_{s}|\geq0.5$ and ${<}5\%$ of simulations return $P_{r_{s}}>0.05$. A correlation that satisfies one of these criteria is considered tentative, and we reject any correlation if neither is satisfied or $P_{r_{s}} > 95\%$.
We apply the Spearman rank correlation test to our full sample and subsamples: Centaurs/JFCs/Transition Objects; KBOs; dwarf planets ($H\leq3$); and non-dwarf planet objects from the KBO, Centaur, JFC, and Transition Object populations ($H>3$). 
Figure \ref{ATLASCorrelations} shows plots of the tested ATLAS parameters for the full object sample, the Centaurs, JFCs, Transition Objects, the KBOs, the dwarf planets, and the non-dwarf planet objects in our sample.

{Table \ref{table_SpearmaRankCorrelationsATLAS} lists the parameters from our Spearman rank correlation tests. 
We find no strong correlations between $\beta_{c}$ and $H$, $\beta_{o}$ and $H$, and $\beta_{o}$ and $H_{c}-H_{o}$ for the full sample and all sub-populations. 
However, we find a tentative negative correlation between $\beta_{c}$ and $H_{c}-H_{o}$ for the full sample and all sub-populations as we cannot reject the null hypothesis to ${>}95\%$ confidence. A tentative negative correlation between $\beta_{o}$ and $H_{c}-H_{o}$ is also found for the Centaur/JFC/Transition Object and non-dwarf planet populations. We also find a tentative negative correlation between $\beta_{c}-\beta_{o}$ and $H_{c}-H_{o}$ for the full sample, the KBO population, and the non-dwarf planet population, yet no significant correlation between these parameters for the dwarf planet population, and a tentative correlation that can almost be ruled out to $95\%$ confidence for the Centaur/JFC/Transition Objects sub-population. No strong correlation is identified for any tested parameter pair. We also find no significant change in correlation results when applying or correlation search algorithm to parameter values derived from the full phase angle range.}

\begin{deluxetable*}{lccccr}
\tablecaption{Results of correlation investigation using ATLAS data {in phase angle range $0.1\leq\alpha\leq6$ deg}. Spearman rank correlation coefficients $r_{s}$, associated null hypothesis probability values $P_{r_{s}}$, percentage of Monte Carlo simulations where $P_{r_{s}}>0.05$ (i.e. we cannot reject a random distribution to 95\% confidence), and number of datapoints analysed for correlation tests between phase curve parameters using ATLAS data. \label{table_SpearmaRankCorrelationsATLAS}}
\tablewidth{0pt}
\tablehead{
\colhead{Parameters} & \colhead{$r_{s}$} & \colhead{$P_{r_{s}}$}  & \colhead{\% Simulations with $P_{r_{s}}>0.05$} & \colhead{N} & \colhead{Correlation?}
}
{
\startdata
\multicolumn{6}{c}{Full Sample}\\
\hline
$\beta_{c}$ vs $H_{c}$& 0.214  & 0.482 & 99.985& 13 & No\\
$\beta_{o}$ vs $H_{o}$& 0.159  & 0.603 & 99.99& 13 & No\\
$\beta_{c}$ vs $H_{c}-H_{o}$& -0.747 & 0.003 & 13.931& 13 & Tentative\\
$\beta_{o}$ vs $H_{c}-H_{o}$& -0.297 & 0.325 & 95.694& 13 & No\\
$\beta_{c}-\beta_{o}$ vs $H_{c}-H_{o}$& -0.648 & 0.017 & 52.726& 13 & Tentative\\
\hline
\multicolumn{6}{c}{Centaurs/JFCs/Transition Objects}\\
\hline
$\beta_{c}$ vs $H_{c}$& 0.257  & 0.623 & 100.0& 6  & No\\
$\beta_{o}$ vs $H_{o}$& -0.314 & 0.544 & 99.792& 6  & No\\
$\beta_{c}$ vs $H_{c}-H_{o}$& -0.771 & 0.072 & 63.171& 6  & Tentative\\
$\beta_{o}$ vs $H_{c}-H_{o}$& -0.657 & 0.156 & 73.967& 6  & Tentative\\
$\beta_{c}-\beta_{o}$ vs $H_{c}-H_{o}$& -0.771 & 0.072 & 94.894& 6  & Tentative\\
\hline
\multicolumn{6}{c}{KBOs}\\
\hline
$\beta_{c}$ vs $H_{c}$& -0.286 & 0.535 & 99.94& 7  & No\\
$\beta_{o}$ vs $H_{o}$& 0.643  & 0.119 & 99.538& 7  & No\\
$\beta_{c}$ vs $H_{c}-H_{o}$& -0.75  & 0.052 & 67.321& 7  & Tentative\\
$\beta_{o}$ vs $H_{c}-H_{o}$& -0.036 & 0.939 & 99.975& 7  & No\\
$\beta_{c}-\beta_{o}$ vs $H_{c}-H_{o}$& -0.786 & 0.036 & 72.76& 7  & Tentative\\
\hline
\multicolumn{6}{c}{Dwarf Planets}\\
\hline
$\beta_{c}$ vs $H_{c}$& 0.143  & 0.787 & 99.3& 6  & No\\
$\beta_{o}$ vs $H_{o}$& 0.771  & 0.072 & 98.216& 6  & No\\
$\beta_{c}$ vs $H_{c}-H_{o}$& -0.714 & 0.111 & 68.087& 6  & Tentative\\
$\beta_{o}$ vs $H_{c}-H_{o}$& -0.086 & 0.872 & 99.969& 6  & No\\
$\beta_{c}-\beta_{o}$ vs $H_{c}-H_{o}$& -0.714 & 0.111 & 97.497& 6  & No\\
\hline
\multicolumn{6}{c}{Non-Dwarf Planets}\\
\hline
$\beta_{c}$ vs $H_{c}$& 0.536  & 0.215 & 100.0& 7  & No\\
$\beta_{o}$ vs $H_{o}$& -0.071 & 0.879 & 97.804& 7  & No\\
$\beta_{c}$ vs $H_{c}-H_{o}$& -0.857 & 0.014 & 43.9& 7  & Tentative\\
$\beta_{o}$ vs $H_{c}-H_{o}$& -0.679 & 0.094 & 79.644& 7  & Tentative\\
$\beta_{c}-\beta_{o}$ vs $H_{c}-H_{o}$& -0.857 & 0.014 & 82.876& 7  & Tentative\\
\enddata
}
\end{deluxetable*}

We also apply our algorithm to the data of \citet{2018MNRAS.481.1848A}, who reported phase coefficients and absolute magnitudes for 114 objects in Johnson-Cousins $V$ and $R$ filters, which exhibit overlap in wavelength range with ATLAS $c$ and $o$. We select only objects from their study with $\beta$ values consistent with positive to within $2\sigma$ and divide their object sample according to our definitions. Table \ref{table_SpearmaRankCorrelationsAL18} lists the results from applying our algorithm to their data. 
{For $\beta$ and $H$, we find tentative negative correlations for the Centaurs/JFCs/Transition Objects sub-population in both the $V$ and $R$ filters.}
{For $\beta$ and $H_{V}-H_{R}$, we find no correlation in V for any sub-population, yet find a tentative positive correlation for the Centaurs/JFCs/Transition Objects and the dwarf planets sub-population.}
{For $\beta_{V}-\beta_{R}$ and $H_{V}-H_{R}$, we find highly significant negative correlations for the full sample, the KBOs, and the non-dwarf planet sub-populations
}

\begin{deluxetable*}{lccccr}
\tablecaption{Results of correlation investigation using data from \citet{2018MNRAS.481.1848A}. Spearman rank correlation coefficients $r_{s}$, associated null hypothesis probability values $P_{r_{s}}$, percentage of Monte Carlo simulations where $P_{r_{s}}>0.05$ (i.e. we cannot reject a random distribution to 95\% confidence), number of datapoints $N$ analysed for correlation tests between phase curve parameters from \citet{2018MNRAS.481.1848A}, and if the parameters exhibit a correlation. \label{table_SpearmaRankCorrelationsAL18}}
\tablewidth{0pt}
\tablehead{
\colhead{Parameters} & \colhead{$r_{s}$} & \colhead{$P_{r_{s}}$}  & \colhead{\% Simulations with $P_{r_{s}}>0.05$} & \colhead{N} & \colhead{Correlation?}
}
\startdata
\multicolumn{6}{c}{Full Sample}\\
\hline
$\beta_{V}$ vs $H_{V}$& -0.278  & 0.005 & 18.208 & 100 & No\\
$\beta_{R}$ vs $H_{R}$& -0.176 & 0.081 & 77.217 & 99 & No\\
$\beta_{V}$ vs $H_{V}-H_{R}$& -0.219 & 0.037 & 81.36 & 91 & No\\
$\beta_{R}$ vs $H_{V}-H_{R}$& 0.408 & 0.0 & 10.796& 91 & No\\
$\beta_{V}-\beta_{R}$ vs $H_{V}-H_{R}$& -0.747 & 0.0 & 0.133 & 91 & Yes\\
\hline
\multicolumn{6}{c}{Centaurs/JFCs/Transition Objects}\\
\hline
$\beta_{V}$ vs $H_{V}$& -0.508  & 0.026 & 14.087 & 19 & Tentative\\
$\beta_{R}$ vs $H_{R}$& -0.521 & 0.022 & 13.153 & 19 & Tentative\\
$\beta_{V}$ vs $H_{V}-H_{R}$& -0.449 & 0.054 & 90.363 & 19 & No\\
$\beta_{R}$ vs $H_{V}-H_{R}$& -0.573 & 0.010 & 75.758 & 19 & Tentative\\
$\beta_{V}-\beta_{R}$ vs $H_{V}-H_{R}$& -0.307 & 0.201 & 84.616 & 19 & No\\
\hline
\multicolumn{6}{c}{KBOs}\\
\hline
$\beta_{V}$ vs $H_{V}$& -0.152  & 0.177 & 99.297 & 81 & No\\
$\beta_{R}$ vs $H_{R}$& -0.110 & 0.330 & 99.959 & 80 & No\\
$\beta_{V}$ vs $H_{V}-H_{R}$& -0.271 & 0.022 & 75.219 & 72 & No\\
$\beta_{R}$ vs $H_{V}-H_{R}$& 0.418 & 0.0 & 19.681 & 72 & No\\
$\beta_{V}-\beta_{R}$ vs $H_{V}-H_{R}$& -0.802 & 0.0 & 0.248 & 72 & Yes\\
\hline
\multicolumn{6}{c}{Dwarf Planets}\\
\hline
$\beta_{V}$ vs $H_{V}$& 0.143  & 0.787 & 100.0 & 6 & No\\
$\beta_{R}$ vs $H_{R}$& 0.6 & 0.208 & 99.623 & 6 & No\\
$\beta_{V}$ vs $H_{V}-H_{R}$& 0.429 & 0.397 & 86.778 & 6 & No\\
$\beta_{R}$ vs $H_{V}-H_{R}$& 0.771 & 0.072 & 86.046 & 6 & Tentative\\
$\beta_{V}-\beta_{R}$ vs $H_{V}-H_{R}$& -0.714 & 0.111 & 96.087 &  6 & No\\
\hline
\multicolumn{6}{c}{Non-Dwarf Planets}\\
\hline
$\beta_{V}$ vs $H_{V}$& -0.241  & 0.019 & 56.606 & 94 & No\\
$\beta_{R}$ vs $H_{R}$& -0.225 & 0.031 & 35.844 & 93 & No\\
$\beta_{V}$ vs $H_{V}-H_{R}$& -0.230 & 0.034 & 81.014 & 85 & No\\
$\beta_{R}$ vs $H_{V}-H_{R}$& 0.377 & 0.0 & 19.221 & 85 & No\\
$\beta_{V}-\beta_{R}$ vs $H_{V}-H_{R}$& -0.745 & 0.0 & 0.28 & 85 & Yes\\
\enddata
\end{deluxetable*}

{We find a highly significant negative correlation using data from \citet{2018MNRAS.481.1848A} for objects with phase coefficient values consistent with positive to within $2\sigma$. We also find similar negative correlations with ATLAS data for the full sample, the KBOs, and the non-dwarf planets sub-populations. However, these correlations are tentative, and an inherently uncorrelated distribution cannot be ruled out to $95\%$ confidence.}
{Jackknife re-sampling of the \citet{2018MNRAS.481.1848A} sample, using sub-sample sizes of $N = 13$ (for the full ATLAS sample) and $N = 7$ (for both the KBO and non-dwarf planet sub-populations), yields no correlation for 16.3\% of simulations. Thus the difference in size between our sample and that of \citet{2018MNRAS.481.1848A} is likely responsible for this discrepancy in correlation certainty. Furthermore, our ATLAS datasets usually have many more observations than those of \citet{2018MNRAS.481.1848A}, and are taken in filters of different wavelength ranges. We therefore {consider our findings to be consistent with those of \citet{2018MNRAS.481.1848A} and } attribute {any} difference {in correlation certainty} to be due to differing sample sizes, dataset sizes, and filter wavelength ranges.  
An increase in the number of objects with phase coefficient and absolute magnitude measurements from datasets comparable to those afforded by ATLAS would allow us 
to potentially recover this negative correlation between relative phase coefficient - the difference in phase coefficient value between broadband filters of differing wavelength range -  and color index.}

\citet{2018MNRAS.481.1848A} reported the strong negative correlation between relative phase coefficient and color index as persisting even when their sample was divided by absolute magnitude and semimajor axis. However, their sample was divided by semimajor axis and absolute magnitude at $a=40$ au and $H=4.5$, in contrast to our thresholds of $a=30$ au and $H=3$.
{Dividing our sample by the sub-population definitions of \citet{2018MNRAS.481.1848A} using $H$ values calculated from ATLAS data, we find tentative correlations for the $a<40$ au and $H>4.5$ subpopulations, and no correlation for the $H<4.5$ au subpopulation, with the $a>40$ au population being too small for an accurate rank statistic.}
Applying our algorithm to their sample divided by their definitions, we also find strong negative correlations between relative phase coefficient and color index for the $H>4.5$ population and $a>40$ au population, and tentative negative correlations for the $a<40$ au population and $H<4.5$ au populations.
{Dividing their sample by our definitions of sub-populations, we find no correlation for Centaurs/JFCs/Transition Objects or the dwarf planets sub-populations, yet a strong correlation for the KBOs and the non-dwarf planet sub-population which is dominated in number by KBOs. This difference in observed correlation between populations may point to a difference in substructure between the surfaces of these objects, likely caused by an evolutionary effect due to differing insolation.}
An increased sample size similar to those of \citet{2016A&A...586A.155A}, \citet{2018MNRAS.481.1848A} and \citet{2019MNRAS.488.3035A}, with precisions of parameter values comparable to those afforded by the ATLAS datasets, would help {to better recover and study this negative correlation between relative phase coefficient and color index, which may in fact be a universal property across different populations of small Solar System objects \citep{2022A&A...667A..81A}, and confirm the existence of this possible distinction.}

\subsection{Cometary Activity Search}

We searched for instances of cometary activity in the ATLAS datasets for all 18 objects in our sample.
Most of our sample show no sign of cometary activity. A small number of objects, namely {
Chiron, 
Echeclus, 
Pluto, 
and 
Quaoar} 
exhibit `bright nights' as defined in Section \ref{CometActivityCorr}, suggesting potential cometary outbursts or long-term activity. 
Table \ref{table_CometaryActivity} lists the bright observations per object identified by our algorithm as potentially indicative of cometary activity, either in the form of short outbursts or longer-term sustained brightening. 
{We discuss each object below}:
\newline

\begin{figure}
\centering
\includegraphics[width=0.75\columnwidth]{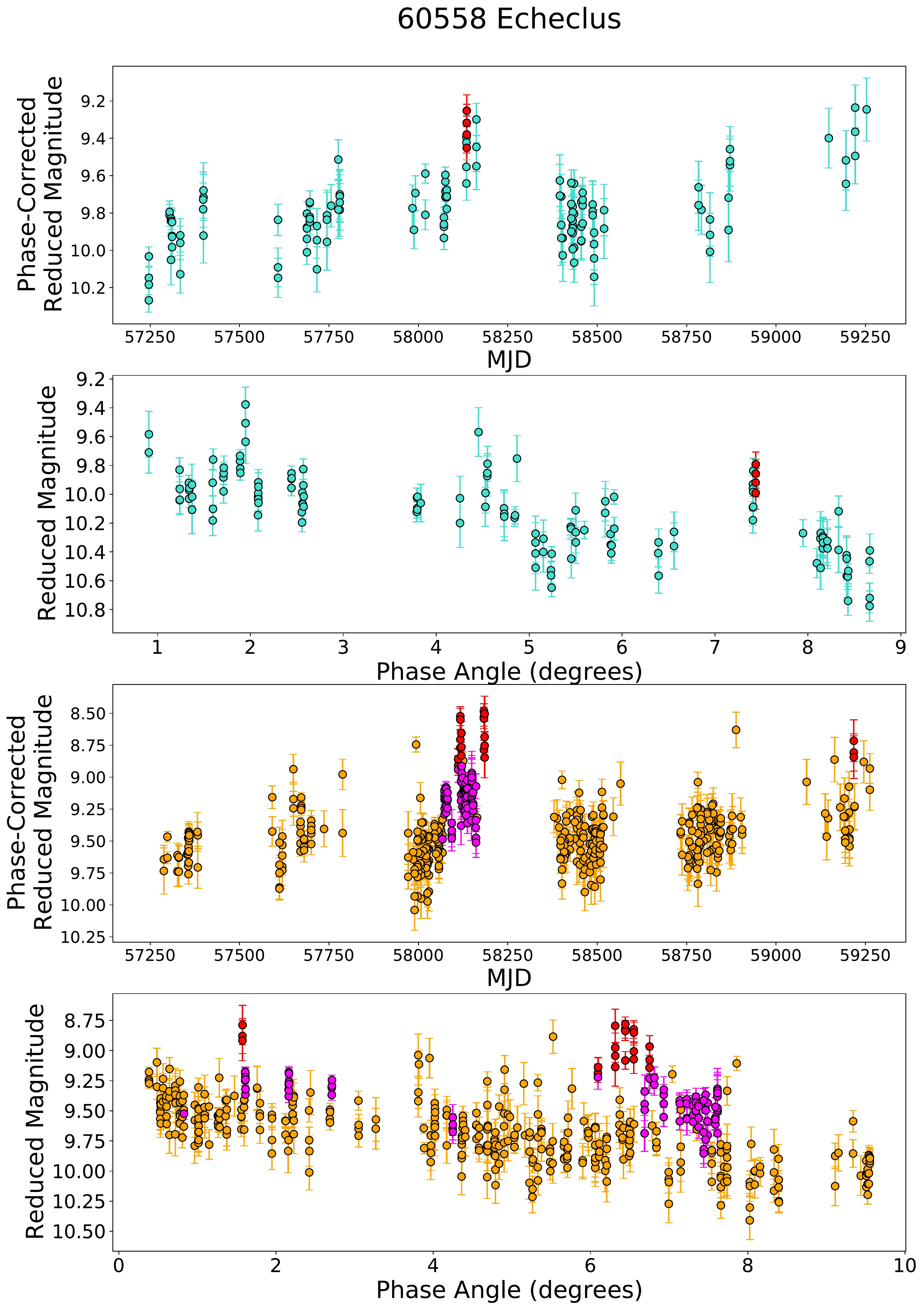}
\caption{Plots of phase-angle corrected reduced magnitude vs. MJD in ATLAS $c$ (turquoise, upper) and $o$ (orange, upper) and phase curves of reduced magnitude vs. phase angle in ATLAS $c$ (turquoise, lower) and $o$ (orange, lower) for (60558) Echeclus. Red points denote observations taken from nights with ${\geq}3$ observations of which ${75\%}$ are brighter than $2\sigma$ from the object's sigma-clipped median magnitude. Magenta points denote observations within a 30 day bin whose median magnitude lies 0.2 magnitudes and 1 standard deviation brighter than the object's sigma-clipped median magnitude.}
\label{CometaryActivityEcheclus}
\end{figure}

\begin{figure}
\centering
\includegraphics[width=0.9\textwidth]{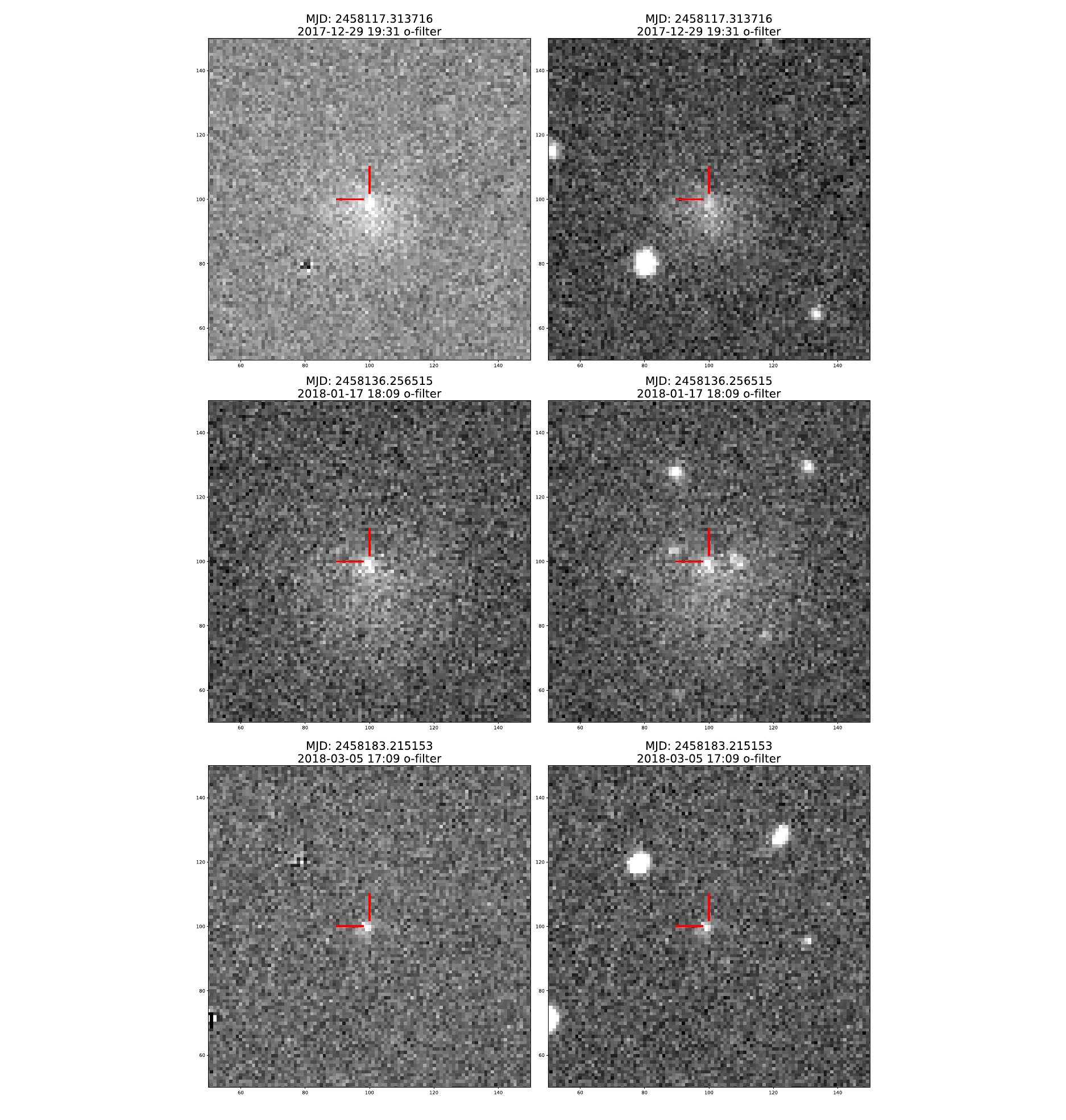}
\caption{
ATLAS difference (left) and reduced (right) images of Echeclus during its observed cometary outburst. Red crosshairs denote Echeclus's position on the images. \emph{Upper:} 2017 December 29 19:31 UTC (MJD 58117.313716) ATLAS $o$ filter images, with fan-shaped coma clearly visible (detected from {our} short outburst search algorithm). \emph{Middle:} 2018 January 17 18:09 UTC (MJD 58136.256515) ATLAS $o$ filter images, with fan-shaped coma clearly visible (detected from our longer-term lower-level activity search algorithm). \emph{Lower:} 2018 March 5 UTC 17:09 (MJD 58183.215153) ATLAS $o$ filter images, showing Echeclus to appear fuzzy compared to background stars, indicating residual coma (detected from our short outburst search algorithm). 
}
\label{OutburstEcheclus}
\end{figure}

\begin{figure}
\centering
\includegraphics[width=0.92\textwidth]{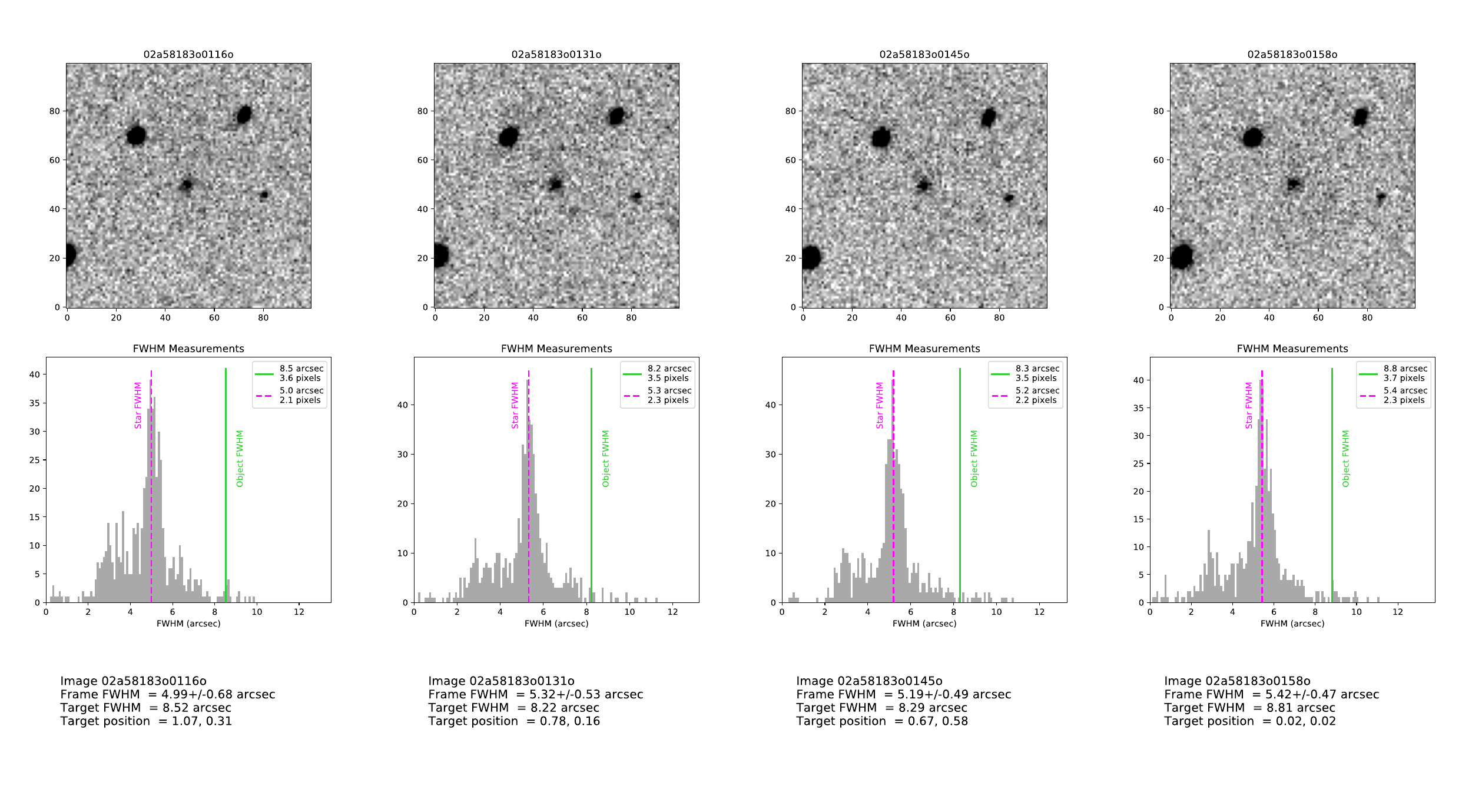}
\caption{PSF FWHM analysis for 2018 March 6 (MJD 58183) ATLAS images of (60558) Echeclus. \emph{Upper row:} 100x100 pixel inset of 1000x1000 pixel reduced ATLAS images centered on Echeclus. \emph{Lower row:} Histograms of FWHM of all background stars in each 1000x1000 pixel ATLAS image, with the median stellar FWHM (dashed magenta vertical line) and Echeclus FWHM (solid green vertical line) plotted for reference. In all images, Echeclus displays significant PSF extension.}
\label{FWHMEcheclus}
\end{figure}

\begin{figure}
\centering
\includegraphics[width=0.65\textwidth]{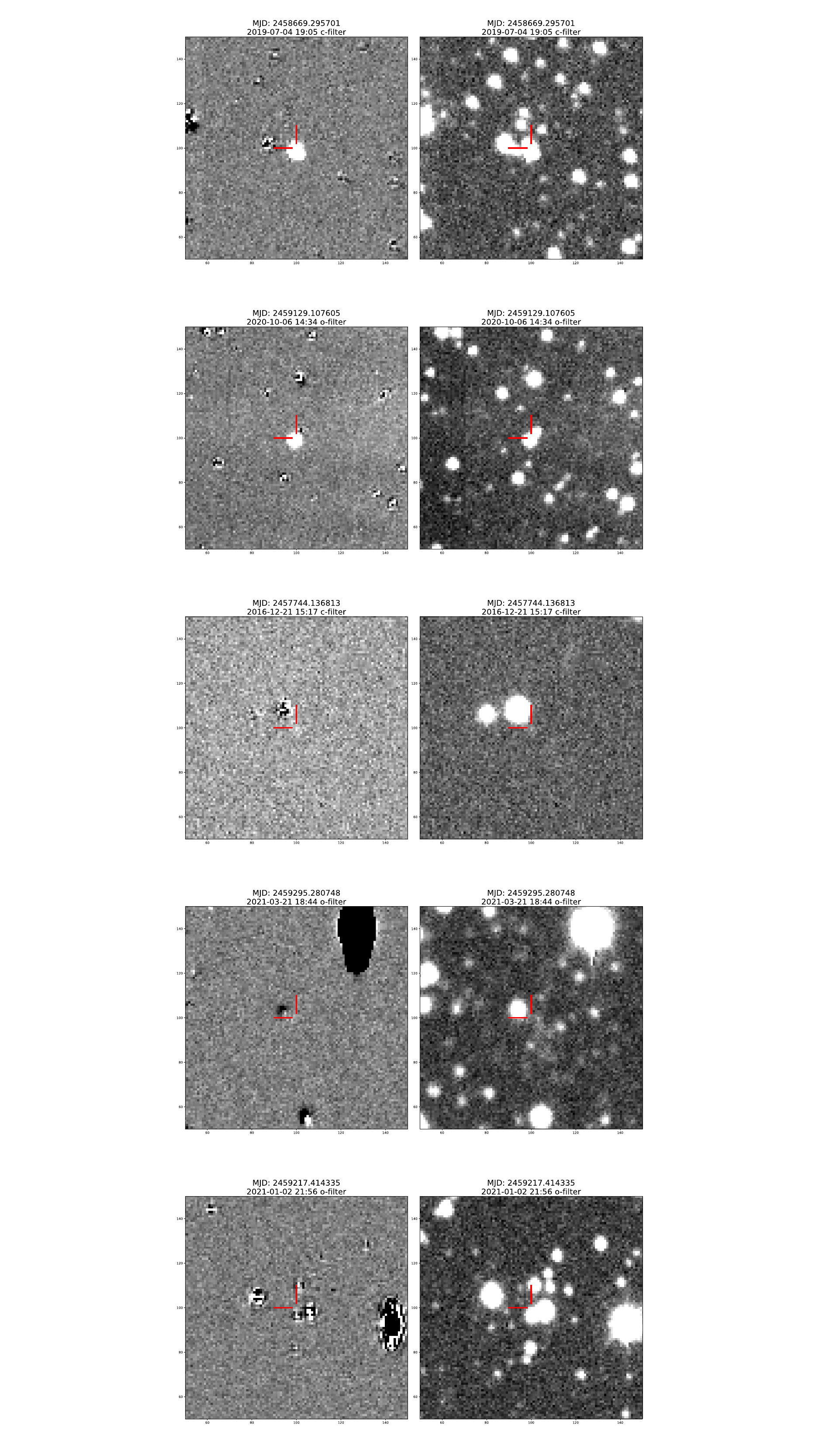}
\caption{Collage of ATLAS difference (left) and reduced (right) images of, in descending order, 
Pluto (2019-07-04 19:05 UTC, $c$ filter),
Pluto (2020-10-06 14:34 UTC, $o$ filter), 
Chiron (2016-12-21 15:17 UTC, $c$ filter), 
Quaoar (2021-03-21 18:44 UTC, $o$ filter),
and Echeclus (2021-01-02 21:56 UTC, $o$ filter). Red crosshairs denote positions of objects in images. In all images, the objects lie close to often poorly-subtracted background stars.}
\label{FalsePositives}
\end{figure}

\emph{Echeclus} - we detect three separate instances of bright nights across the ATLAS baseline, as shown in Figure \ref{CometaryActivityEcheclus}. The first, seen in the $c$ filter on 2018 January 17 (MJD 58135) and the $o$ filter from 2017 December 24 (MJD 58111) to 2018 January 2 (MJD 58120), coincides in time with a previously reported 2017 December outburst \citep{2018JBAA..128...51J,2019AJ....158..255K}. During this outburst, Echeclus exhibited a fan-shaped coma which remained clearly visible in the ATLAS images for approximately 1 month, as seen in Figure \ref{OutburstEcheclus}. This activity seemed to build to a second outburst observed as two consecutive bright nights, 2018 March 6 (MJD 58183) and 2018 March 8 (MJD 58185). Analysis of the full width half maximum (FWHM) of the PSF of Echeclus during the night of 2018 March 6 (the night during which the PSF of Echeclus was more clearly visible) reveals significant spread compared to those of background stars, as shown in Figure \ref{FWHMEcheclus}. The PSF FWHM of Echeclus compared to that of the background stars for this date are listed in Table \ref{table_EcheclusFWHM}, indicative of a visible coma.
The third instance appears as a single bright night in the $o$ filter on 2018 April 9, however analysis of the ATLAS images reveals Echeclus to lie close to a poorly-subtracted background star during this time, as seen in Figure \ref{FalsePositives}, making this a result of background stellar flux contamination, and not a comet-like outburst.

\begin{figure}
\centering
\includegraphics[width=0.75\columnwidth]{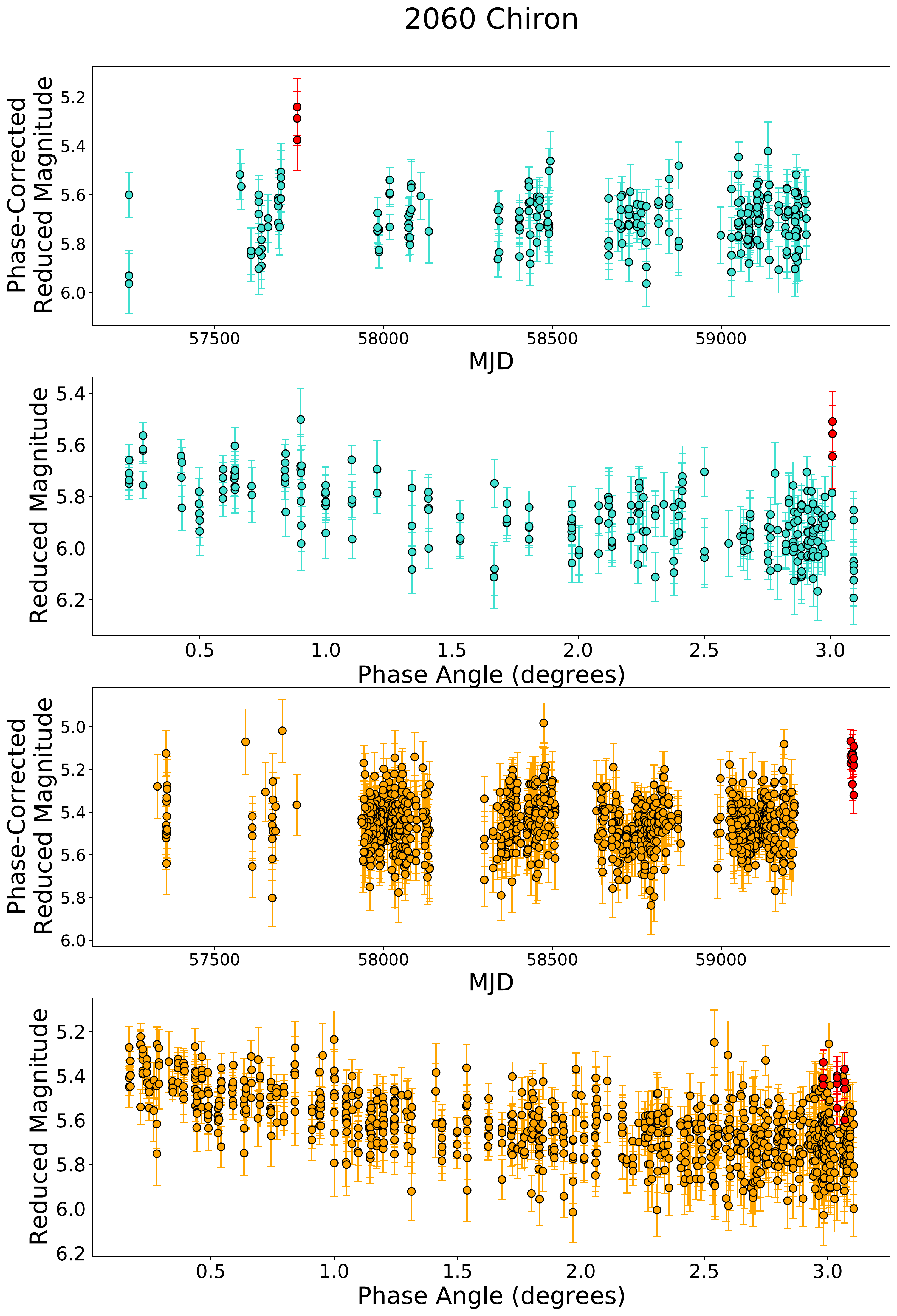}
\caption{Plots of phase-angle corrected reduced magnitude vs. MJD in ATLAS $c$ (turquoise, upper) and $o$ (orange, upper) and phase curves of reduced magnitude vs. phase angle in ATLAS $c$ (turquoise, lower) and $o$ (orange, lower) for (2060) Chiron. Red points denote observations taken from nights with ${\geq}3$ observations of which ${75\%}$ are brighter than $2\sigma$ from the object's median sigma-clipped magnitude.}
\label{CometaryActivityChiron}
\end{figure}

\begin{figure}
\centering
\includegraphics[width=0.9\textwidth]{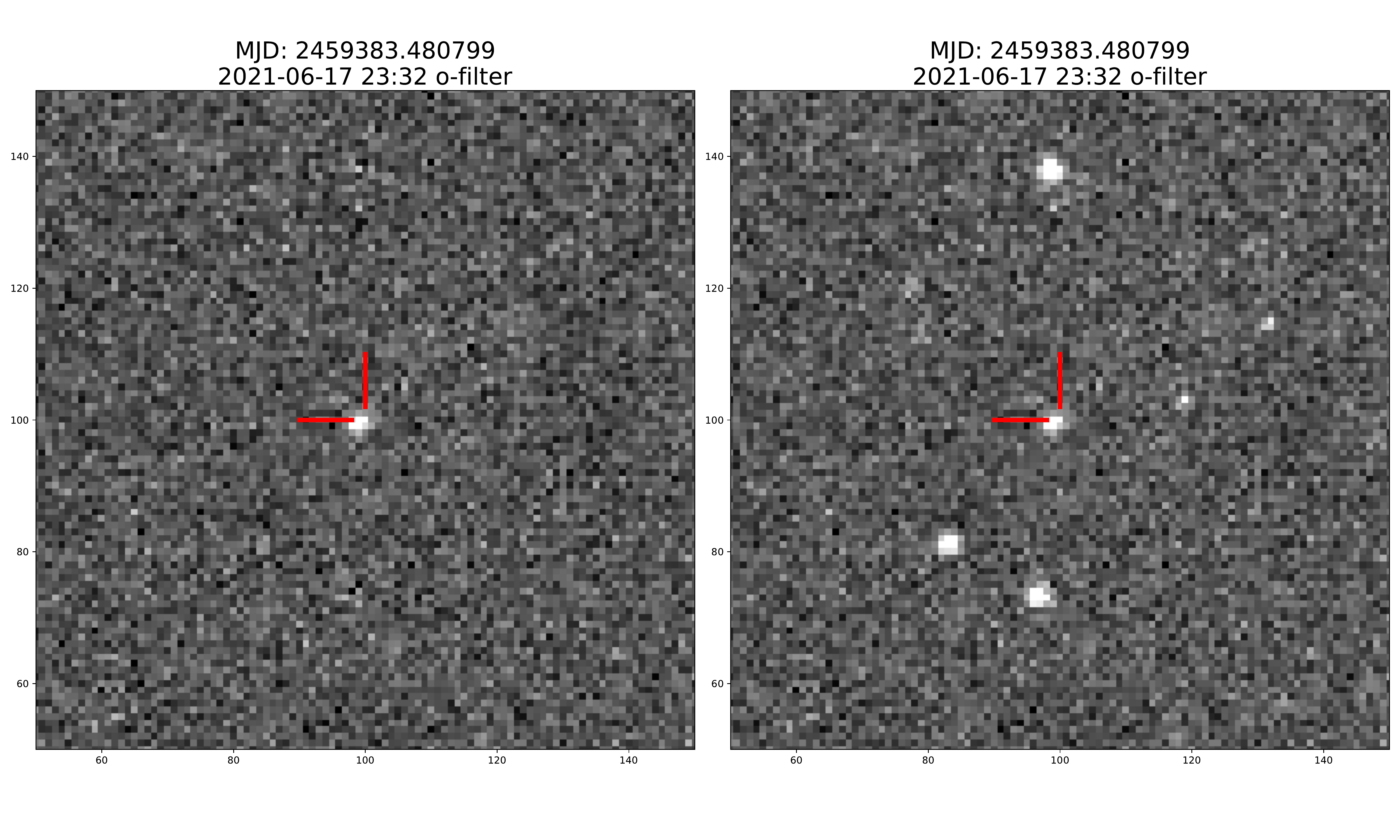}
\caption{2021 June 17 23:32 UTCMJD 59383.480799 ATLAS difference image (left) and reduced image (right) of Chiron during its 2021-2022 active epoch, taken in ATLAS $o$ filter. Red crosshairs denote Chiron's position on the images. Chiron's signal remains point-like from visual inspection.}
\label{OutburstChiron}
\end{figure}

\begin{figure}
\centering
\includegraphics[width=0.92\textwidth]{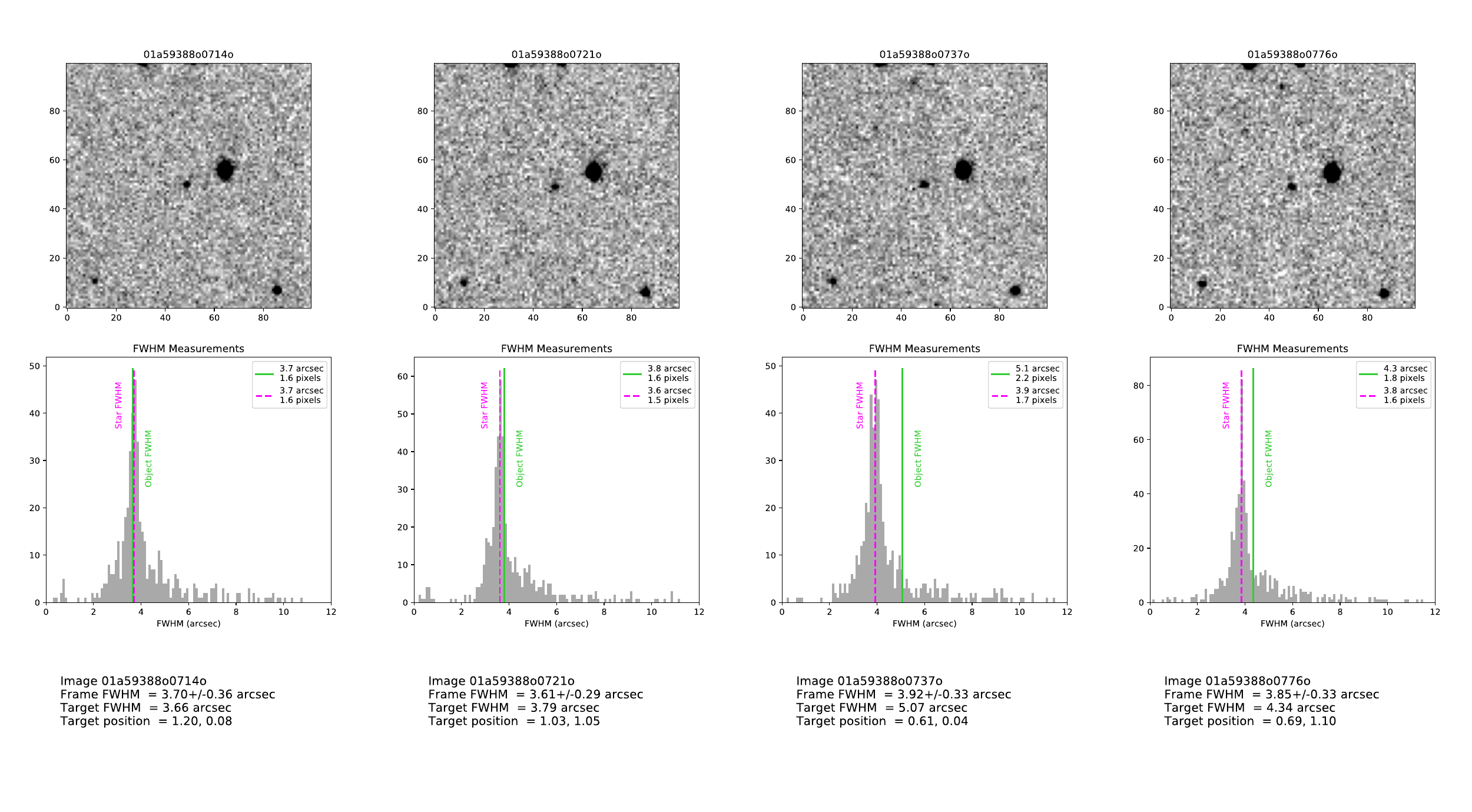}
\caption{PSF FWHM analysis for 2021 June 23 (MJD 59388) ATLAS images of (2060) Chiron. \emph{Upper row:} 100x100 pixel inset of 1000x1000 pixel reduced ATLAS images centered on (2060) Chiron. \emph{Lower row:} Histograms of FWHM of all background stars in each 1000x1000 pixel ATLAS image, with the median stellar FWHM (dashed magenta vertical line) and (2060) Chiron FWHM (solid green vertical line) plotted for reference. In all images, the FWHM of (2060) Chiron remains consistent with the histogram distribution of the FWHM of background stars, with no significant extension.}
\label{FWHMChiron}
\end{figure}

\emph{Chiron} - we detect two separate instances of bright nights, highlighted in Figure \ref{CometaryActivityChiron}. The first is a single bright night at 2016 December 22 (MJD 57744) observed in $c$; visual inspection reveals Chiron lies close to a poorly-subtracted background star, as seen in Figure \ref{FalsePositives}, thus this brightening is due to stellar contamination.
The second instance is an increase in brightness constituting three consecutive bright nights starting at 2021 June 18 (MJD 59383), the first three nights of observation of a new apparition of Chiron. No coma or tail is visible in the ATLAS images, and Chiron's flux signal appears stellar (see Figures \ref{OutburstChiron} and \ref{FWHMChiron} and Table \ref{table_EcheclusFWHM}), yet this brightness increase exceeds Chiron's amplitude of rotational modulation. This new epoch of increased activity is reported in \citet{2021ATel14903....1D} and \citet{2021RNAAS...5..211D}. Further analysis of Chiron is beyond the scope of this study, and will be subject of a later paper.

\begin{figure}
\centering
\includegraphics[width=0.75\columnwidth]{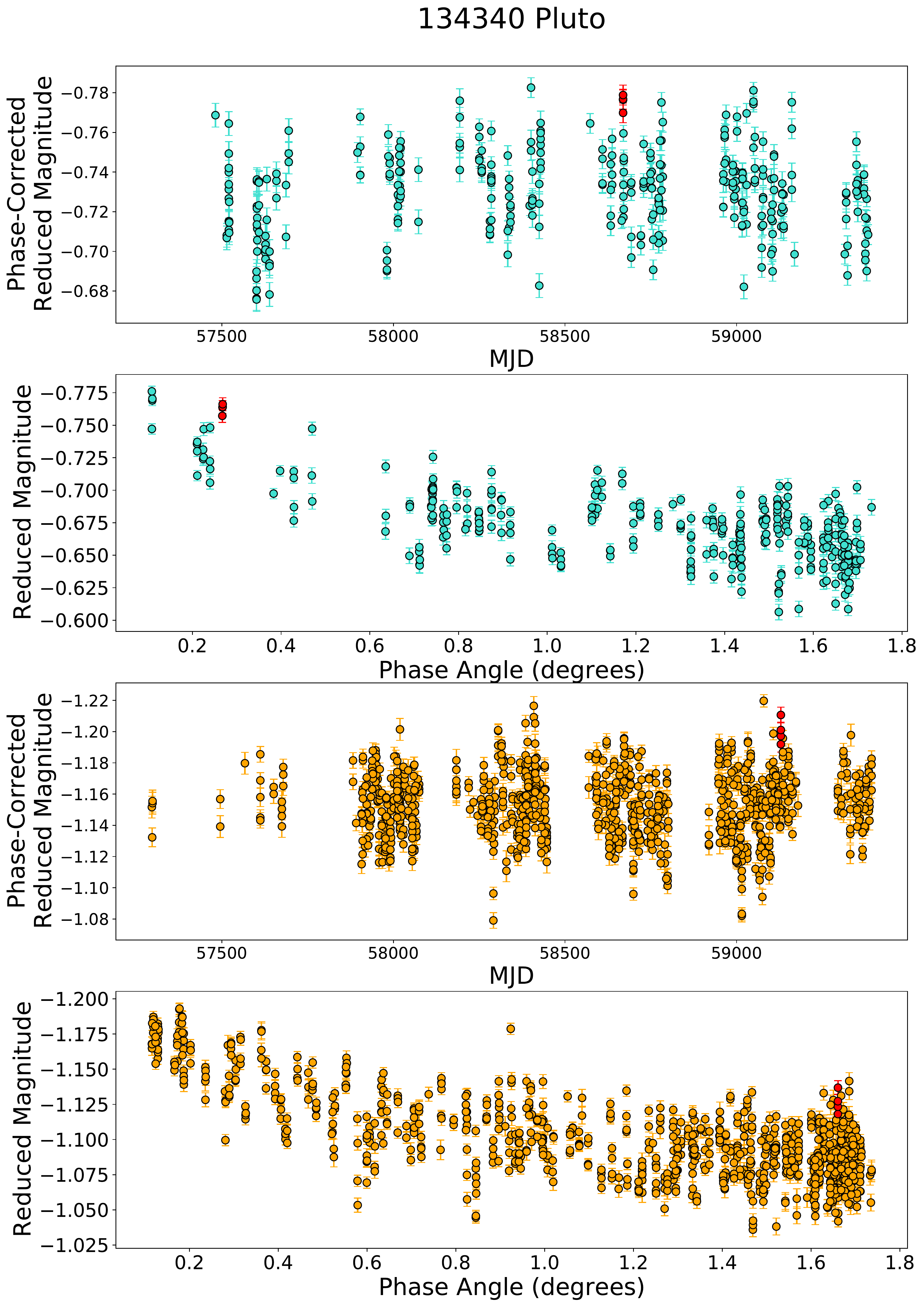}
\caption{Plots of phase-angle corrected reduced magnitude vs. MJD in ATLAS $c$ (turquoise, upper) and $o$ (orange, upper) and phase curves of reduced magnitude vs. phase angle in ATLAS $c$ (turquoise, lower) and $o$ (orange, lower) for (134340) Pluto, also corrected for rotational modulation. Red points denote observations taken from nights with ${\geq}3$ observations of which ${75\%}$ are brighter than $2\sigma$ from the sigma-clipped object's median magnitude.}
\label{CometaryActivityPluto}
\end{figure}

\emph{Pluto} - {we detect two bright nights at 2019 July 5 (MJD 58669) and 2020 October 7 (MJD 59129), highlighted in Figure \ref{CometaryActivityPluto}, yet neither are caused by cometary activity. Visual inspection of ATLAS images of each night shows that Pluto is contaminated by background stars, as seen in Figure \ref{FalsePositives}.}

\begin{figure}
\centering
\includegraphics[width=0.75\columnwidth]{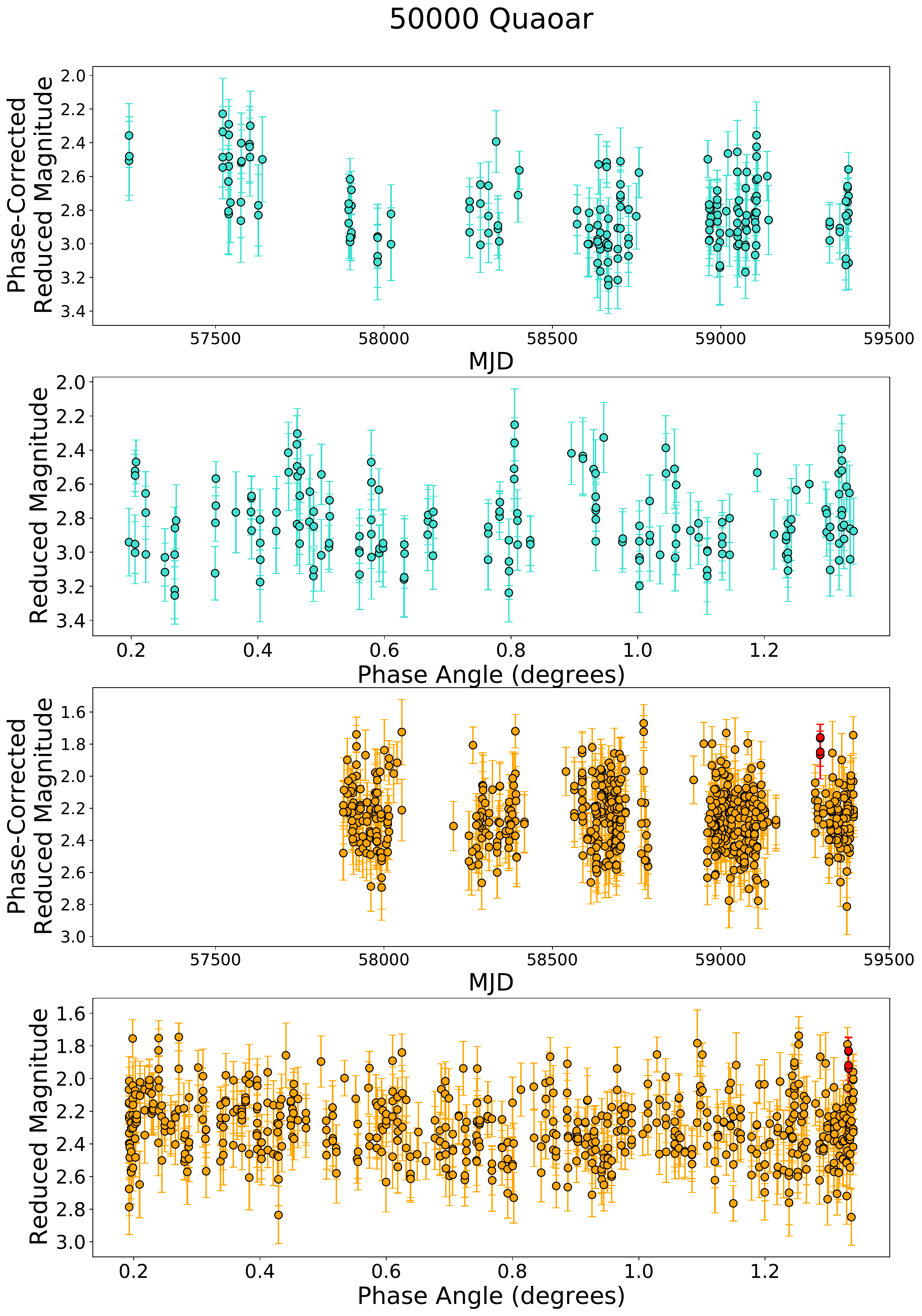}
\caption{Plots of phase-angle corrected reduced magnitude vs. MJD in ATLAS $c$ (turquoise, upper) and $o$ (orange, upper) and phase curves of reduced magnitude vs. phase angle in ATLAS $c$ (turquoise, lower) and $o$ (orange, lower) for (50000) Quaoar, also corrected for rotational modulation. Red points denote observations taken from nights with ${\geq}3$ observations of which ${75\%}$ are brighter than $2\sigma$ from the object's sigma-clipped median magnitude.}
\label{CometaryActivityQuaoar}
\end{figure}

\emph{Quaoar} - we detect a single bright night at 2021 March 22 (MJD 59295), highlighted in Figure \ref{CometaryActivityQuaoar}. We find Quaoar lies close to a poorly-subtracted background star in all observations from this night, as seen in Figure \ref{FalsePositives}, contaminating its photometry. 

We also detect two further instances of brightening for both Echeclus and 2014 QA43 which may correspond to long-term activity, described below:

\emph{Echeclus} - we detect an instance of sustained brightening of 0.2 mag and ${>}1$ standard deviation of the data from the sigma-clipped median ranging from 2017 November 10 (MJD 58067) to 2018 February 12 (MJD 58161), as highlighted in Figure \ref{CometaryActivityEcheclus}. This coincides in time with Echeclus's ${>}2\sigma$ outburst in 2017 December, and a fan-shaped coma is clearly visible in the ATLAS images during this time, as seen in Figure \ref{OutburstEcheclus}. 

\begin{figure}
\centering
\includegraphics[width=0.75\columnwidth]{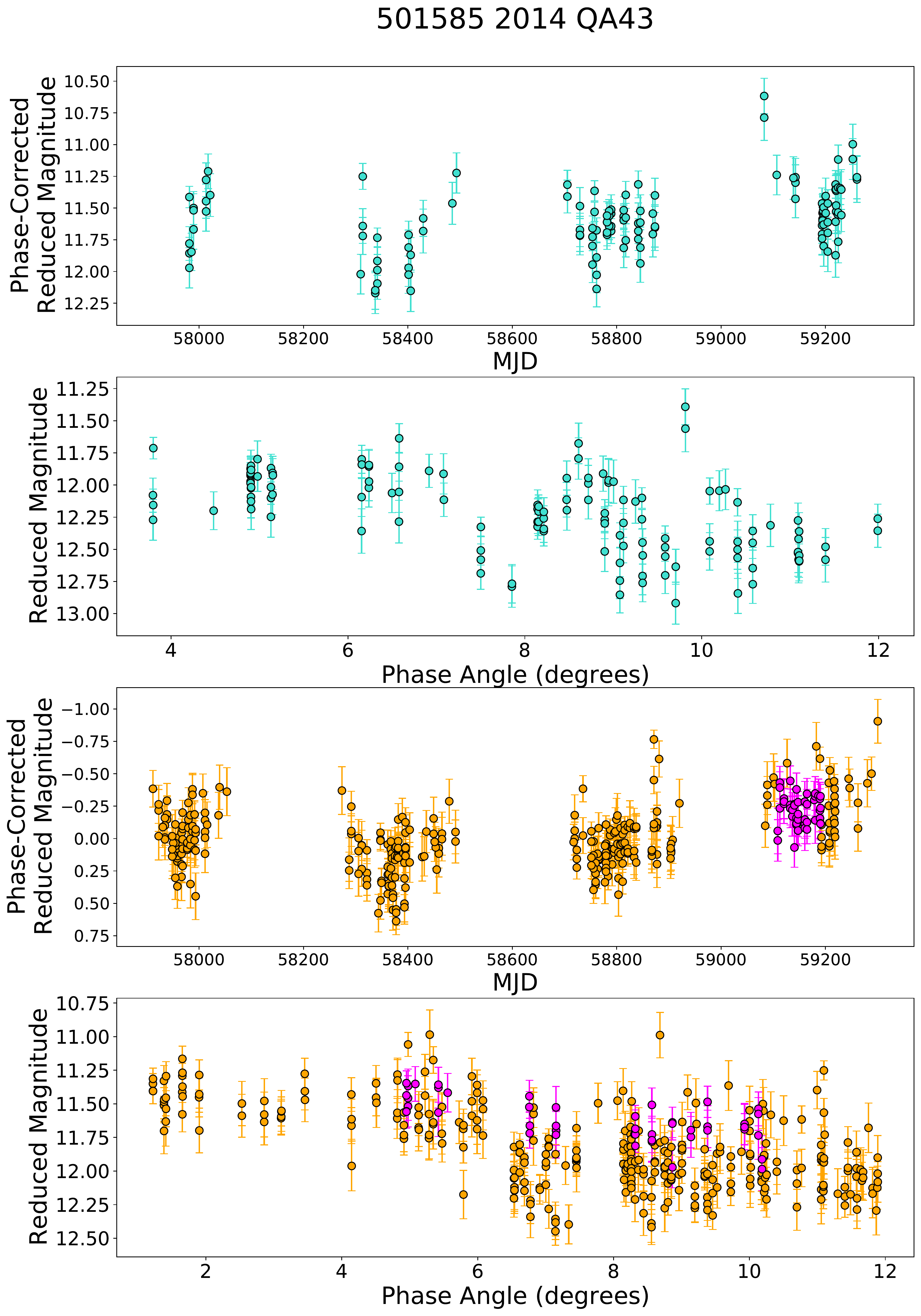}
\caption{Plots of phase-angle corrected reduced magnitude vs. MJD in ATLAS $c$ (turquoise, upper) and $o$ (orange, upper) and phase curves of reduced magnitude vs. phase angle in ATLAS $c$ (turquoise, lower) and $o$ (orange, lower) for (501585) 2014 QA43. Red points denote observations taken from nights with ${\geq}3$ observations of which ${75\%}$ are brighter than $2\sigma$ from the sigma-clipped object's median magnitude. Magenta points denote observations within a 30 day bin whose median magnitude lies 0.2 magnitudes and 1 standard deviation brighter than the object's sigma-clipped median magnitude.}
\label{CometaryActivity2014QA43}
\end{figure}

\begin{figure}
\centering
\includegraphics[width=0.92\textwidth]{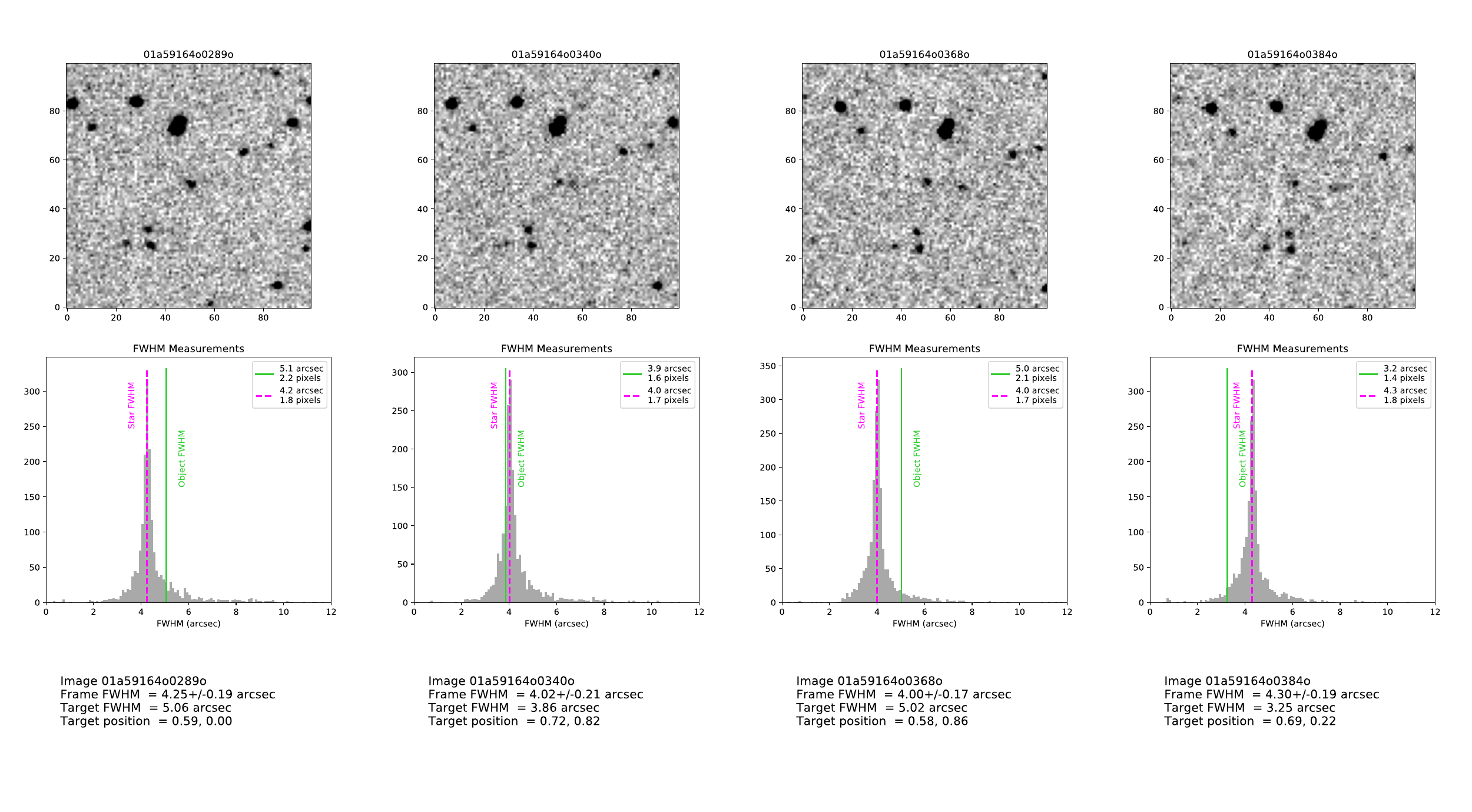}
\caption{PSF FWHM analysis for 2020 November 11 (MJD 59164) ATLAS images of (501585) 2014 QA43. \emph{Upper row:} 100x100 pixel inset of 1000x1000 pixel reduced ATLAS images centered on (501585) 2014 QA43. \emph{Lower row:} Histograms of FWHM of all background stars in each 1000x1000 pixel ATLAS image, with the median stellar FWHM (dashed magenta vertical line) and (501585) 2014 QA43 FWHM (solid green vertical line) plotted for reference. The PSF of (501585) 2014 QA43 remains consistent with the histogram distribution of the FWHM of background stars, with no significant extension.}
\label{FWHM2014QA43}
\end{figure}

\emph{2014 QA43} - we detect an instance of brightening ranging in time from {2020 September 16 (MJD 59108)} to 2020 December 7 (MJD 59190), spanning a large fraction of the observations from that apparition, as shown in Figure \ref{CometaryActivity2014QA43}. The median phase-corrected reduced magnitude of this apparition is brighter than that of previous years of data by approximately 0.3 mag. Measurements of the FWHM of 2014 QA43 reveals no significant broadening compared to background stars, as seen in Figure \ref{FWHM2014QA43}. It remains unclear if this brightening is due to cometary activity not visible in the images or due to the changing aspect angle of the object. 

In conclusion, only Chiron and Echeclus are found to exhibit definite comet-like activity. It is not surprising that we find these objects have been active during the ATLAS survey, as both Chiron \citep[]{1988IAUC.4684....2B,1990AJ....100..913L,1991MNRAS.250..115D,1993Icar..104..234M,1995Natur.373...46E,1997P&SS...45.1607L,2001P&SS...49.1325S,2002Icar..160...44D} and Echeclus \citep[]{2006CBET..563....1C,2006IAUC.8656....2C,2011IAUC.9213....2J,2016MNRAS.462S.432R,2018JBAA..128...51J,2019AJ....157...88S,2019AJ....158..255K} have well-recorded histories of comet-like activity in the literature. Both these objects have been classified by previous studies as belonging to the Centaur population, members of which have been known to exhibit comet-like outbursts. However, unlike comets, Centaur activity is decoupled from perihelion. The mechanism responsible for cometary activity on Centaurs currently remains elusive at present.
\citet{2009AJ....137.4296J} found that from a population of 23 Centaurs, the median perihelion of active Centaurs was smaller than that of inactive ones, and proposed a transition from amorphous to crystalline water ice to be the cause of Centaur activity.
Considering only objects in our sample which satisfy the Centaur definition of \citet{2009AJ....137.4296J}, we find that Echeclus and Chiron have perihelia between those of Jupiter and Saturn, with those of our inactive Centaurs extending to beyond that of Saturn. Though our sample is too small to accurately quantify median perihelia, our observations of Centaur activity appear to be consistent with the findings of \citet{2009AJ....137.4296J}. 

All of the variable (or comet-like activity) observations in the ATLAS dataset that correspond to instances of potential cometary activity were removed from the dataset before phase curve analyses were conducted. {If any found instance of possible cometary activity was extended in time, we remove all datapoints in both filters between the start date and end date of the activity. Furthermore, any data identified by our algorithm as a bright night yet showed background star contamination were removed from the dataset.} We note that the number of observations taken during an outburst remains a limiting factor in detecting such activity.
For example, despite occurring during the ATLAS observation baseline, a previously-recorded outburst of Echeclus in 2016 August \citep{2016MNRAS.462S.432R} did not cause a ${>}2\sigma$ increase in the ATLAS data, and observations were too sparse for it to be detected as lower-level cometary activity. Furthermore, any lower-level cometary activity that persisted for several consecutive apparitions would likely remain undetected, thereby reducing the accuracy of any applied phase curve fit. However, our algorithm has allowed us to detect an instance of cometary activity on Chiron despite displaying no visible coma or comet-like tail. Such activity would remain undetected from analysis of ATLAS postage stamps, yet is readily detectable in our algorithm. Therefore, we present our phase curve fits as the best analysis that can be carried out within the limits of the dataset.

\begin{deluxetable*}{lccccr}
\tablecaption{Instances of nights of observation potentially indicative of cometary activity. \label{table_CometaryActivity}}
\tablewidth{0pt}
\tablehead{
\colhead{Object} & \colhead{Filter} & \colhead{Start Date} & \colhead{End Date} & \colhead{Cometary Activity?}  & \colhead{Notes}
}
\startdata
Echeclus & c & 2018-01-17 (MJD 58135) & 2018-01-17 (MJD 58135) & Short outburst & Fan-shaped coma visible\\
Echeclus & o & 2017-12-24 (MJD 58111) & 2018-01-02 (MJD 58120) & Short outburst & Fan-shaped coma visible\\
Echeclus & o & 2018-03-06 (MJD 58183) & 2018-03-06 (MJD 58183) & Short outburst & PSF extension\\
Echeclus & o & 2021-01-03 (MJD 59217) & 2021-01-03 (MJD 59217) & No & Background stellar flux contamination\\
Echeclus & o & 2017-11-10 (MJD 58067) & 2018-02-12 (MJD 58161) & Sustained brightening & Fan-shaped coma visible\\
Chiron & c & 2016-12-22 (MJD 57744) & 2016-12-22 (MJD 57744) & No & Background stellar flux contamination\\
Chiron & o & 2021-06-18 (MJD 59383) & 2021-06-27 (MJD 59392) & Sustained brightening & No visible activity or PSF extension \\
&&&&& yet brightening beyond literature amplitude \\
&&&&& and spread of data from previous apparitions\\
Pluto & c & 2019-07-04 (MJD 58669) & 2019-07-04 (MJD 58669) & No & Background stellar flux contamination\\
Pluto & o & 2020-10-07 (MJD 59129) & 2020-10-07 (MJD 59129) & No & Background stellar flux contamination\\
Quaoar & o & 2021-03-22 (MJD 59295) & 2021-03-22 (MJD 59295) & No & Background stellar flux contamination\\
2014 QA43 & o & {2020-09-16 (MJD 59108)} & 2020-12-07 (MJD 59190) & Unknown & No visible coma or No PSF extension\\
\enddata
\end{deluxetable*}

\begin{deluxetable*}{lcccr}
\tablecaption{PSF FWHM analysis for selected bright nights of Echeclus, Chiron, and 2014 QA43 compared to PSF FWHM of background stars. \label{table_EcheclusFWHM}}
\tablewidth{0pt}
\tablehead{
\colhead{Object} & \colhead{UTC Date and Time} & \colhead{MJD} & \colhead{Echeclus PSF FWHM} & \colhead{Stellar PSF FWHM}\\
\colhead{} & \colhead{} & \colhead{} & \colhead{(arcsec)}  & \colhead{(arcsec)}
}
\startdata
Echeclus & 2018-03-05 17:18 & 58183.215153 & $8.52$ & $4.99\pm0.68$\\
Echeclus & 2018-03-05 17:20 & 58183.222678 & $8.22$ & $5.32\pm0.53$\\
Echeclus & 2018-03-05 17:30 & 58183.229396 & $8.29$ & $5.19\pm0.49$\\
Echeclus & 2018-03-05 17:39 & 58183.23565 & $8.81$ & $5.42\pm0.47$\\
\hline
Chiron & 2021-06-22 23:07 & 59388.463541 & $3.66$ & $3.70\pm0.36$\\
Chiron & 2021-06-22 23:12 & 59388.466769 & $3.79$ & $3.61\pm0.29$\\
Chiron & 2021-06-22 23:22 & 59388.474141 & $5.07$ & $3.92\pm0.33$\\
Chiron & 2021-06-22 23:48 & 59388.492191 & $4.34$ & $3.85\pm0.33$\\
\hline
2014 QA43 & 2020-11-11 00:32 & 59164.522729 & $5.06$ & $4.25\pm0.19$\\
2014 QA43 & 2020-11-11 01:07 & 59164.546531 & $3.86$ & $4.02\pm0.21$\\
2014 QA43 & 2020-11-11 02:12 & 59164.592126 & $5.02$ & $4.00\pm0.17$\\
2014 QA43 & 2020-11-11 02:23 & 59164.599976 & $3.25$ & $4.30\pm0.19$\\
\enddata
\end{deluxetable*}

\subsection{$HG_{1}G_{2}$ Model}

\begin{figure}
\centering
\includegraphics[width=0.95\textwidth]{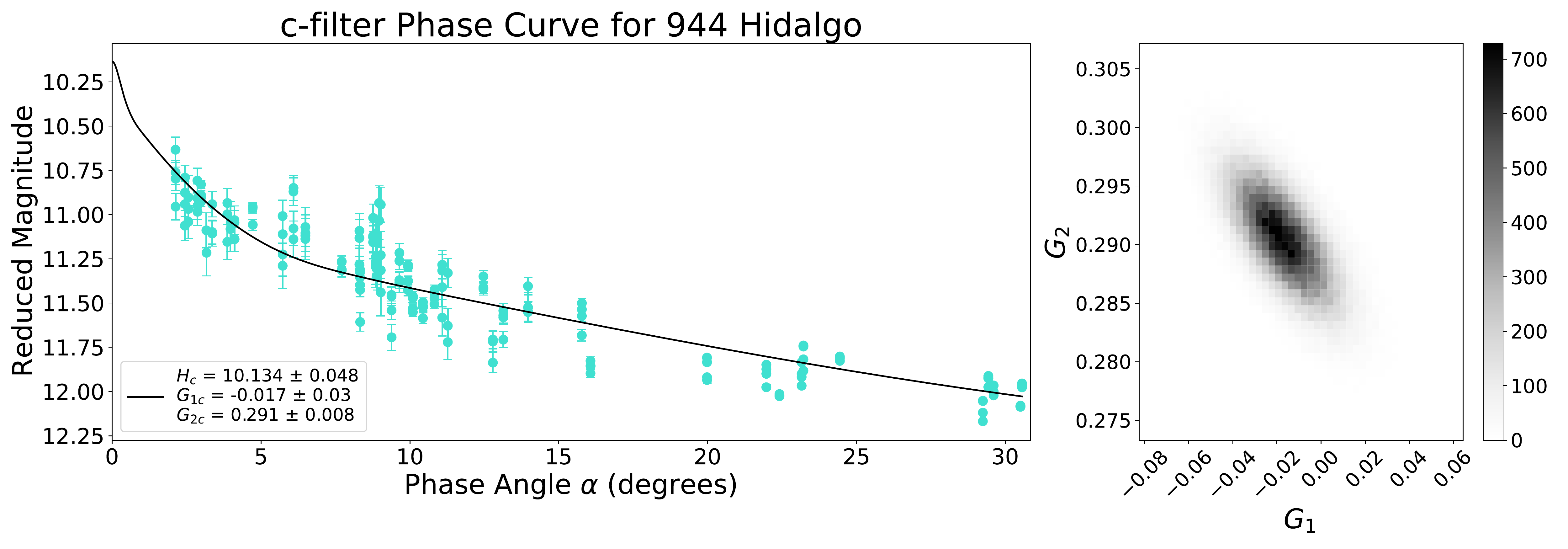}
\caption{\emph{Left plot:} $HG_{1}G_{2}$ model fit to \emph{c} filter
    phase curve of (944) Hidalgo. \emph{Right plot:} 2D histogram of
    Monte Carlo solutions to slope parameters $G_{1}$ and $G_{2}$ of
    $HG_{1}G_{2}$ fit. {Best fit function extended to $\alpha = 0$
    deg for ease of viewing the extrapolated absolute magnitude value
    and any potential opposition surge.}
    The complete figure set (4 images) is available in the online published version of the paper.}
\label{PhCHG1G2Hidalgoc}
\end{figure}

\begin{figure}
\centering
\includegraphics[width=16cm,height=21cm]{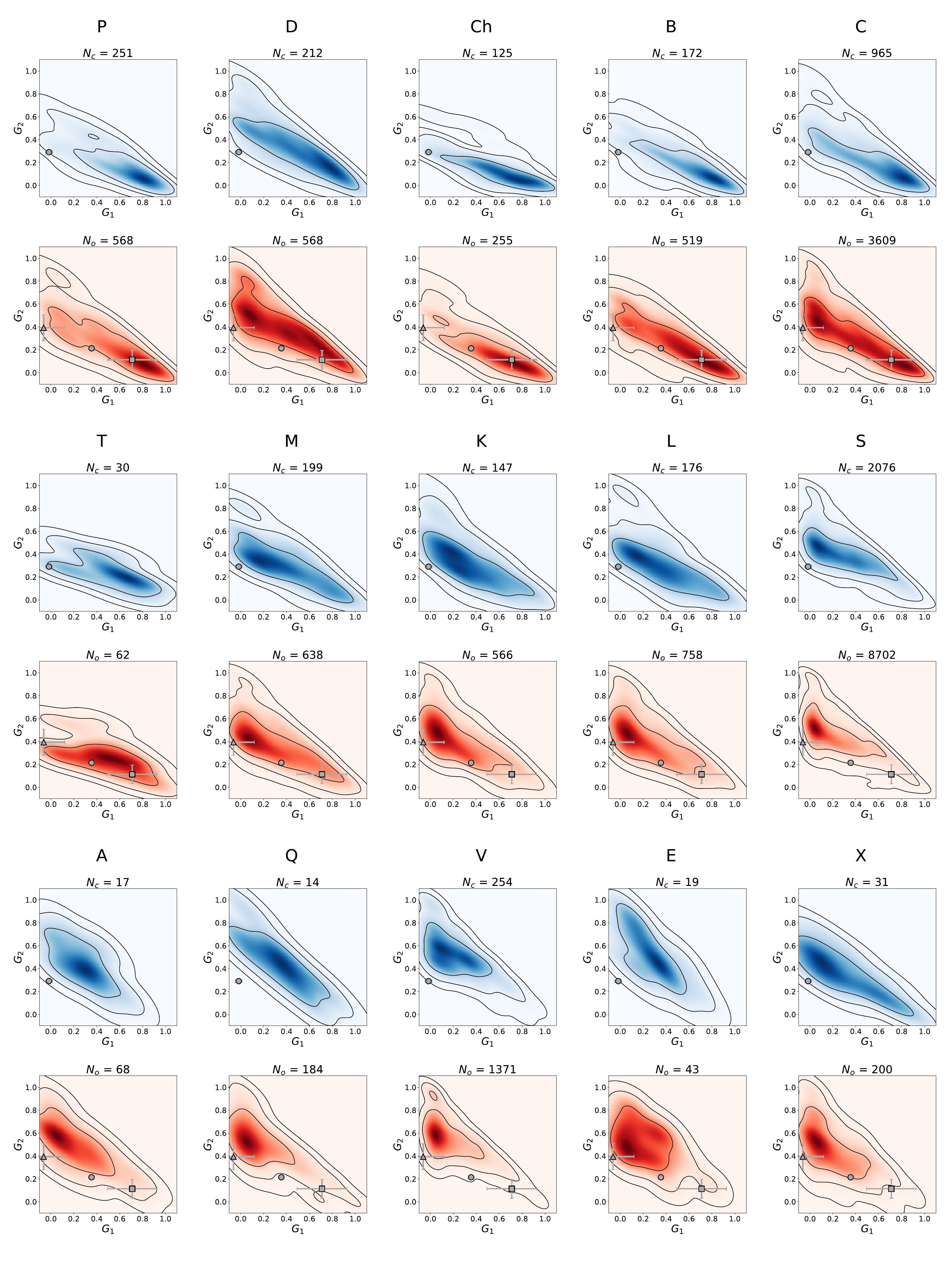}
\caption{$G_1$,$G_2$ KDE distributions of taxonomic complexes as mapped by \citet{2021Icar..35414094M} in $c$ and $o$ ATLAS filters, with $G_1$ and $G_2$ parameters of Hidalgo (circle), Narcissus (square), and 2014 QA43 (triangle) phase curves overplotted. Contours mark KDE levels at which 68.3\% ($1\sigma$), 95\% ($2\sigma$) and 99.7\% ($3\sigma$) of values are encompassed. Uncertainties for Hidalgo datapoints are smaller than the plot symbol.}
\label{G1G2Plots}
\end{figure}

We apply the $HG_{1}G_{2}$ phase curve model to the $c$ filter
    dataset of Hidalgo and the $o$ filter phase curve of Hidalgo,
    Narcissus and 2014 QA43. The resulting fits to the data are shown
    in Figure \ref{PhCHG1G2Hidalgoc} and its online Figure set. Table
    \ref{table_HG1G2} lists the absolute magnitudes $H$ and slope
    parameters $G_1$ and $G_2$ of these phase curves.
Our best-fit slope parameters 
are consistent within their uncertainties with monotonic functions of positive slope, as predicted from surface reflectance models \citep[]{2010Icar..209..542M,2021Icar..35414094M}. \citet{2021Icar..35414094M} utilized serendipitous ATLAS observations of main belt asteroids to fit their phase curves with the $HG_{1}G_{2}$ function to determine possible correlations of slope parameters with main belt asteroid (MBA) taxonomic class. We aim to compare our best-fit slope parameters with those from the taxonomic classes to infer the surface types of Hidalgo, Narcissus and 2014 QA43.
Figure \ref{G1G2Plots} displays the kernel density estimation (KDE) plots of $G_1$ and $G_2$ values of 19708 MBAs as measured by \citet{2021Icar..35414094M} sorted by taxonomic complex, with our derived slope parameters for our three JFCs overplotted with their associated error bars. We include the $c$ filter slope {parameter of Hidalgo} for completeness, but for analysis we consider only the slope parameters for our objects in the $o$ filter as they were derived from more observations. As seen in Figure \ref{G1G2Plots}, 
the $o$ filter slope parameters of all three fitted JFCs are consistent within uncertainties with the KDE distributions of all asteroid taxonomic complexes to $3\sigma$, with greater correspondences to the carbonaceous C-type complexes. 
The surfaces of Narcissus and 2014 QA43 remain unclassified, while Hidalgo has been previously classified as a D-type asteroid \citep[]{1984PhDT.........3T,2017Icar..284...30B,2022A&A...658A.158G}. Our $G_{1}$ and $G_{2}$ parameters for Hidalgo are consistent with its previous classification, lying within the 95\% confidence interval, albeit {just} outside the core of the distribution of the D-type complex.

\begin{deluxetable*}{lcccr}
\tablecaption{$H$, $G_1$, and $G_2$ parameters, their associated $2\sigma$ uncertainties, and the number of datapoints $N$ for objects that satisfy criteria for fitting with the $HG_{1}G_{2}$ model.\label{table_HG1G2}}
\tablewidth{0pt}
\tablehead{
\colhead{Object} & \colhead{$H$} & \colhead{$G_1$}  & \colhead{$G_2$} & \colhead{N}
}
\startdata
Hidalgo (c filter)      & $10.135 \pm 0.048$ & $-0.017 \pm 0.03$ & $0.291 \pm 0.008$ & 178 \\
Hidalgo (o filter)      & $10.193 \pm 0.026$ & $0.355 \pm 0.024$ & $0.214 \pm 0.006$ & 730 \\
Narcissus (o filter)  & $13.044 \pm 0.124$ & $0.71 \pm 0.216$ & $0.113 \pm 0.083$ & 204 \\
2014 QA43 (o filter) & $10.899 \pm 0.093$ & $-0.063 \pm 0.178$ & $0.393 \pm 0.106$  & 270
\enddata
\end{deluxetable*}

\hfill
\hfill
\hfill

\section{Conclusions}

We use serendipitous observations from the long baseline and high cadence ATLAS survey to generate phase curves for 18 small Solar System objects, including 7 KBOs, 4 Centaurs, and 5 JFCs. We have measured the linear phase coefficients for all 18 objects in both wide-band $c$ and $o$ ATLAS filters. The results are as follows: 

\begin{enumerate}
    
    \item Negative phase coefficients $\beta$ values noted in the literature become more positive with the larger datasets from ATLAS, and are thus an artifact of small datasets with large scatter, and not a real physical phenomenon. 

    \item {We search for correlations between phase curve parameters using phase coefficient and absolute magnitude values re-measured across the phase angle range $0.1\leq\alpha\leq6$ deg. W}e find no correlation between the linear phase coefficients and absolute magnitude for either ATLAS filter. We report no correlation between phase coefficient and color index in the $o$ filter, and a tentative negative correlation in the $c$ filter, consistent with the results of \citet{2018MNRAS.481.1848A} and \citet{2019MNRAS.488.3035A}. {We find 
a tentative negative correlation between relative phase coefficient and object color index in the ATLAS $c$ and $o$ filters, 
consistent with the findings of \citet{2018MNRAS.481.1848A} and \citet{2019MNRAS.488.3035A}, though an increase in sample size would help to better resolve this correlation.}

    \item 
    Splitting our sample into sub-populations, no significant correlation is found between phase coefficient and absolute magnitude in either filter. 
Tentative correlations are found in the ATLAS $c$ filter between phase coefficient and color index for all sub-populations, and in the $o$ filter for the Centaur/JFCs/Transition Objects and non-dwarf planet sub-populations. 
    We find a tentative negative correlation between relative phase coefficient and object surface color for the KBO population, the dwarf planet population, and all non-dwarf planet objects, consistent with the results of \citet{2018MNRAS.481.1848A} and \citet{2019MNRAS.488.3035A}. 
    The apparent lack of correlation exhibited by the Centaurs/JFCs/Transition Objects compared to the KBOs may indicate a difference in surface properties, but a larger sample size is needed to confirm this.

    \item We detect comet-like activity exhibited by Echeclus (corresponding to previous literature outbursts) and Chiron (corresponding to a newly discovered epoch of activity reported by \citet{2021RNAAS...5..211D}){, and we also detect potential cometary activity for 2014 QA43.} This complements established methods looking for extended PSFs and faint coma or tails in images to detect epochs of cometary activity.

    \item We apply the $HG_{1}G_{2}$ model to the phase curves of 3 JFCs - Hidalgo, Narcissus, and 2014 QA43. We find their best-fitting slope parameters to be more consistent with carbonaceous main belt asteroid taxonomic complexes than silicaceous ones, indicating low-albedo, carbon-rich surfaces for these objects.

\end{enumerate}

The long baseline and high observation cadence of ATLAS has yielded larger photometric datasets per object in our sample than any previous study.
Future surveys utilising larger telescope diameters, such as the {Rubin Observatory} Legacy Survey of Space and Time (LSST) by the Vera C. Rubin Observatory \citep[]{2009arXiv0912.0201L,2019arXiv190108549J,2019ApJ...873..111I}, though operating with a lower observation cadence than ATLAS, will help reduce uncertainties in magnitude measurements, aiding discovery of and correction for rotational {lightcurves}. 
LSST will also significantly increase the number of objects whose phase curves can be analysed by probing to fainter limiting magnitudes, allowing increased accuracy for characterizing larger numbers of the KBO, Centaur, and JFC populations.

\begin{acknowledgements}

\section{Acknowledgements}

This work has made use of data from the Asteroid Terrestrial- impact Last Alert System (ATLAS) project. The Asteroid Terrestrial- impact Last Alert System (ATLAS) project is primarily funded to search for near earth asteroids through NASA grants NN12AR55G, 80NSSC18K0284, and 80NSSC18K1575; byproducts of the NEO search include images and catalogs from the survey area. The ATLAS science products have been made possible through the contributions of the University of Hawaii Institute for Astronomy, the Queen’s University Belfast, the Space Telescope Science Institute, the South African Astronomical Observatory, and the Millennium Institute of Astrophysics (MAS), Chile. MES and MMD were supported by the UK Science Technology Facilities Council (STFC) grants ST/V000691/1 and ST/V506990/1, respectively. LJS acknowledges support by the European Research Council (ERC) under the European Union’s Horizon 2020 research and innovation program (ERC Advanced Grant KILONOVA No. 885281).

This research has made use of data and services provided by the Horizons system of the Jet Propulsion Laboratory, and has made use of services provided by NASA's Astrophysics Data System. All photometric data used in this study is available from the online ATLAS Forced Photometry Server at \url{https://fallingstar-data.com/forcedphot/}. 

The authors thank the researchers Grigori Fedorets, David Jackson, Lucy Dolan, and Will Bate for useful feedback on the manuscript. The authors thank the entire ATLAS Collaboration for their productive discussion on this paper. {The authors thank the anonymous reviewers for their feedback that improved the manuscript of this paper.}

\facilities{ATLAS (Maunaloa and Haleakalā telescopes)}

\end{acknowledgements}

\software{Astropy 
\citep[]{2013A&A...558A..33A,2018AJ....156..123A},
Jupyter Notebook \citep{soton403913},
lmfit \citep{2014zndo.....11813N},
math \citep{van1995python},
Matplotlib \citep{Hunter:2007},
Numpy \citep{2011CSE....13b..22V,harris2020array},
os \citep{van1995python},
Pandas \citep{reback2020pandas},
python (\url{https://www.python.org}),
SAOImageDS9 \citep{2019zndo...2530958J},
sbpy \citep{2019JOSS....4.1426M},
SciPy \citep{2020NatMe..17..261V},
seaborn \citep{2021JOSS....6.3021W}}

\bibliography{new.ms.bib}{}
\bibliographystyle{aasjournal.bst}

\end{document}